\documentclass{article}
\usepackage{graphicx}

\begin{document}

\title{Time delay between the optical and X-ray outbursts in the high mass X-ray transient A0535+26/HDE245770}

\author{F. Giovannelli\thanks{INAF - Istituto di Astrofisica e Planetologia Spaziali -
Area di Ricerca di Tor Vergata - Via del Fosso del Cavaliere 100, I
00133 Roma, Italy}\,\,, G.S. Bisnovatyi-Kogan\thanks{Space Research Institute, Profsoyuznaya 84/32,
Moscow 117810, Russia}\,\,, A.S. Klepnev\thanks{Space Research Institute, Profsoyuznaya 84/32,
Moscow 117810, Russia}}

\date{}
\maketitle

\begin{abstract}
The optical behaviour of the Be star in the high mass X-ray transient A0535+26/HDE245770  shows that at the periastron typically there is an enhancement  in the luminosity of order 0.02 to few tenths mag, and  the X-ray outburst happens about 8 days after the periastron. We construct a  quantitative model of this event, basing on the  a nonstationary accretion disk behavior, connected with a high ellipticity of the orbital motion.  The ephemeris used in this paper --  JD$_{\rm opt-outb}$ = JD$_0$(2,444,944) $\pm$ n(111.0 $\pm$ 0.4) days are derived from the orbital period of the system  P$_{\rm orb} = 111.0 \pm 0.4$ days, determined by Priedhorsky \& Terrell (1983), and from the optical flare of December 5, 1981 (Giovannelli et al., 1985) (here after 811205-E; E stands for the Event occurred at that date) that triggered the subsequent X-ray outburst of December 13, 1981 (Nagase et al., 1982)  (here after 811213-E).
We explain the observed time delay between  the peaks of the optical and X-ray outbursts in this system by the time of radial motion of the matter in the accretion disk, after an increase of the mass flux in the vicinity of a periastral point in the binary. This time is determined by the turbulent viscosity, with the parameter $\alpha=0.1-0.3$.
The increase of the mass flux is a sort of flush that reaches the external part of the accretion disk around the neutron star, producing
an enhancement in the optical luminosity. The consequent X-ray flare happens when the matter reaches the hot central parts of the accretion disk, and the neutron star surface.
\end{abstract}

\section{Introduction}

The X-ray source A 0535+26 was discovered by Ariel V satellite on 14 April, 1975 (Coe et al., 1975). The X-ray source was in outburst with the intensity of $\approx 2$ Crab), and showed a pulsation at $\sim 104$ s (Rosenberg et al., 1975). The hard X-ray spectrum during the decay from the April 1975 outburst became softer, being the 19 May spectrum with $E^{-0.8}$, and the 1 June with    $E^{-1.1}$ (Ricketts et al., 1975).

Between 13 and 19 April, 1975, as the nova brightens, the spectra show some evidence of steepening. The best fit of the experimental data between roughly 27 and 28 April was compatible with 8 keV black-body curve (Coe et al., 1975). The X-ray source decayed from the outburst with an {\it e}-folding time of 19 days in the energy range 3-6 keV (Kaluzienski et al., 1975). The Be star HDE 245770 was discovered as the optical counterpart of A 0535+26 by Bartolini et al. (1978), and was classified as O9.7IIIe star by Giangrande et al. (1980). This is a robust classification, still resistant to many attacks.

Complete reviews of this system can be found in the papers
by Giovannelli et al. (1985), Giovannelli \& Sabau-Graziati
(1992) -- here after GSG92 --, and Burger et al. (1996).

Briefly, the properties of this systems, placed at distance of $1.8 \pm 0.6$ kpc (Giangrande et al., 1980), can be summarized as follows: hard X-ray transient, long period X-ray pulsar -- the secondary star -- orbiting around the primary O9.7IIIe star. The masses are of $\sim 1.5 \pm 0.3$ M$_\odot$ (Joss \& Rappaport 1984; Thorsett et al. 1993; van Kerkwijk, van Paradijs, J. \& Zuiderwijk, 1995), and 15 M$_\odot$ (Giangrande et al., 1980) for the secondary and primary stars, respectively. The eccentricity is e = 0.497 (Finger et al., 1994). Usually the primary star does not fill its Roche lobe (de Loore et al., 1984).

The trigger for writing this paper has been the results reported in the paper by Giovannelli \& Sabau-Graziati (2011) where they
emphasize the discovery of low-energy indicators of high-energy processes. These indicators are UBVRI magnitudes
and Balmer lines of the optical companion. Particular unusual activity of the primary star -- usually at the periastron
passage of the neutron star -- indicates that an X-ray flare is drawing near. The shape and intensity of X-ray outbursts
are dependent on the strength of the activity of the primary.

 By using two measurements in optical during two identical decays from relative maxima of the luminosity of HDE 245770, Bartolini et al. (1983) determined the orbital period of the system HDE 245770/A 0535+26 as P$_{\rm orb} = 110.856 \pm 0.002$ days. They assumed the time of the maximum flare luminosity observed by R\"{o}ssiger (1978) as reference maximum, according with JD(L$_{max}) = 2,443,496.37$ + n $\times$ 110.856.
Thus, Bartolini et al. (1983) obtained for the maximum of the optical flare of December 5, 1981 (JD 2,444,944.5) (Giovannelli et al., 1985) (here after 811205-E; E stands for the Event) the computed time JD 2,444,937.5 not in contrast with the observed time. For this reason Bartolini et al. (1983) gave JD $2,444,944 \pm 10$ as time of the occurrence of the 811205-E.

But, since the 811205-E is clearly peaked at that date and triggered the subsequent short X-ray outburst of December 13, 1981 (811213-E) (Nagase et al., 1982), Giovannelli \& Sabau-Graziati (2011) assumed the ephemeris of the system as JD$_{\rm opt-outb}$ = JD$_0$(2,444,944) $\pm$ n(110.856 $\pm$ 0.002) days.

Thus they concluded that the passage of the neutron star at the periastron occurs with a periodicity of 110.856 $\pm$ 0.002 days and the various kinds of X-ray outbursts of A0535+26 --
following the definitions reported in the review by GSG92 -- occur just after $\sim$ 8 days. The intensity of X-ray outbursts range from $\approx 0.1-8$ Crab depending on the state of the primary companion O9.7 IIIe star: i) `normal outbursts' ($\approx 0.1-0.5$) Crab when the O9.7 IIIe star is `quiescent' (steady stellar wind); ii) anomalous or noisy outbursts' ($\approx 0.5-1$ Crab) when the O9.7IIIe star is in a `turbulent' state; some puffs of material, superimposed to the steady stellar wind, are expelled; iii) `casual or giant outbursts' ($\approx 1-8$ Crab) when the O9.7IIIe star is `very active' (i.e. expelling a shell). At the periastron the optical luminosity increases from $\approx 0.02$ to $\approx 0.2$ mag from case i) to iii).

The time delay between optical and X-ray outbursts, starting from 811205-E is becoming longer for the most recent outbursts. This suggests that the orbital period determined by Bartolini et al. (1983) and used by Giovannelli \& Sabau-Graziati (2011) is slightly too short.

Thus, in this paper we assume the orbital period determined by Priedhorsky \& Terrell (1983) by using X-ray data: P$_{\rm orb} = 111.0 \pm 0.4$ days, and the ephemeris JD$_{\rm opt-outb}$ = JD$_0$(2,444,944) $\pm$ n(111.0 $\pm$ 0.4) days. The 111-day orbital period is completely in agreement with the many determinations reported in the literature
(from optical data, e.g. Guarnieri et al., 1985; de Martino et al., 1985; Hutchings, 1984; Janot-Pacheco, Motch \& Mouchet, 1987. From X-ray data, e.g. Nagase et al., 1982; Priedhorsky \& Terrell, 1983; Motch et al., 1991; Finger et al., 1996; Coe et al., 2006).

However, for the purposes of this paper the orbital period of the system is not so crucial.

Moreover, the suggestion of the possible presence of a temporary accretion disk around the X-ray pulsar when approaches the periastron (Giovannelli \& Zi\'{o}kowski, 1990), that was confirmed by X-ray measurements of Finger et al. (1996), and deeply discussed by Giovannelli et al. (2007) when the presence of a temporary accretion disk was detected by means of doubling in the He I emission lines, gave us a hint for trying to quantitatively describe the origin of the 8 days time delay between the optical and X-ray outburst in A 0535+26/HDE 245770 (Flavia' star) system.

 In the following section the reader will find a panorama of experimental results enough complete coming from optical and X-ray measurements of the X-ray/Be system A 0535+26/HDE 245770. Only sometimes X-ray and optical measurements were taken in the same epoch, sometimes only optical data were obtained and other times only X-ray data were available. Thus, the reader would pay attention in looking at the various experimental situations, but all having the common denominator: the epoch of the periastron passage of the neutron star around the Be star is always before X-ray outbursts of $\sim 8$ days. If optical data are available, at the periastron passage the optical luminosity shows a relative maximum or the beginning of a decay from it. If X-ray data are available, the X-ray outburst follows the periastron passage of $\sim 8$ days. If both optical and X-ray data are available, optical and X-ray outbursts are separated by $\sim 8$ days.

\section{Some relevant experimental results}

In the following we will present several experimental facts that clearly demonstrate the delay between the relative enhancement of the optical luminosity of HDE 245770, occurring at the periastron passage, and the consequent X-ray outburst of the X-ray pulsar A0535+26. We can divide the experimental facts in three sets: (i) optical and X-ray data are available around the periastron passage; (ii) optical data are available without X-ray data around the periastron passage; (iii) X-ray data are available, but no optical data are available around the periastron passage. In all the three cases we have plotted a line in correspondence with the periastron passage.  In all the cases, the X-ray outbursts follow the epoch of the periastron passage of $\sim 8$ days.

 Of course, if simultaneous optical and X-ray measurements would be always available, our claim about the delay between optical and X-ray outbursts should be further supported.

Figure 1 shows the optical relative maximum luminosity in U, B, and V bands of HDE 245770 occurred on December 5, 1981 (811205-E) (Giovannelli et al., 1985), and the date (December 13, 1981) when the subsequent X-ray outburst occurred (811213-E) (Nagase et al., 1982). 811205-E   is used for providing the ephemeris of the system (JD$_{\rm opt-outb}$ = JD$_0$(2,444,944) $\pm$ n(111.0 $\pm$ 0.4) days) used in this paper.

%%%%%%%%%%%%%%%%%%% FIGURE 1 %%%%%%%%%%%%%%%%%
\begin{figure}%[!hbp]%[!ht]%[!hbp]%[h][h!]
\begin{center}
\includegraphics[width=4.3cm]{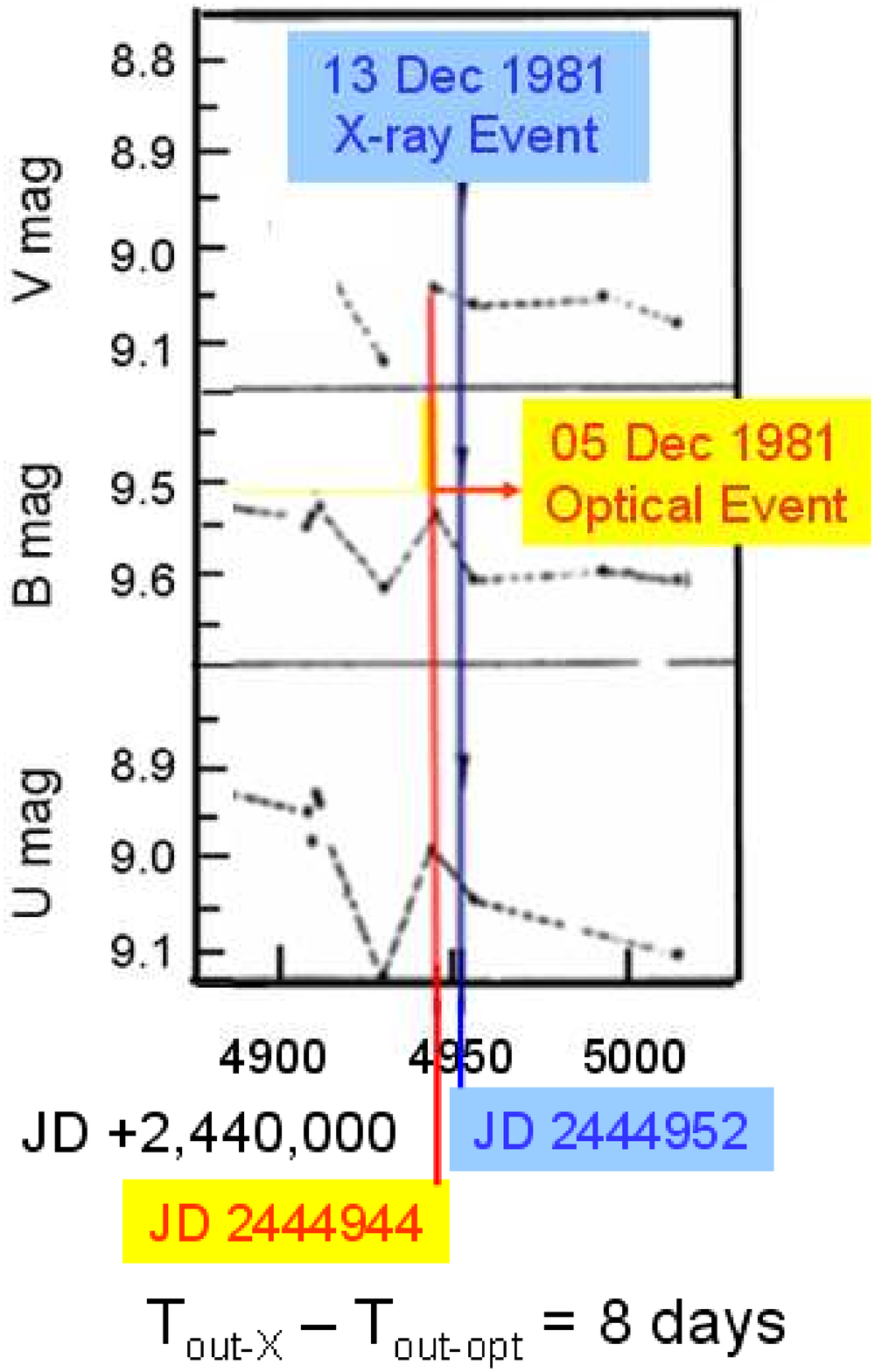}\\
\caption{Optical outburst on December 5, 1981 (811205-E) (Giovannelli et al., 1985)
  marked with a red line  occurred
8 days before the short and sharp X-ray outburst of December 13,
1981 (811213-E) (Nagase et al., 1982)  marked with a blue line.} \label{lc1}
\end{center}
\end{figure}
%%%%%%%%%%%%%%%%%%%%%%%%%%%%%%%%%%%%%%%%%%%%%%

Five cycles after 811205-E, optical data in V band (Gnedin et al., 1988) and X-ray data from Solar Maximum Mission (SMM) in the range 20--163 keV (Sembay et al., 1990), reported in GSG92, are available. Figure 2 reports those data. It is remarkable to note that the X-ray peak reached on June 18--20, 1983 (JD 2,445,504 -- 506) occurs $\sim 5-7$ days after the periastron passage where the V luminosity is going down from a relative maximum.  This fact will be commented later in this section after the description of Fig. 9.

%%%%%%%%%%%%%%%%%%% FIGURE 2 %%%%%%%%%%%%%%%%%
\begin{figure}%[!hbp]%[!ht]%[!hbp]%[h][h!]
\begin{center}
\includegraphics[width=6cm]{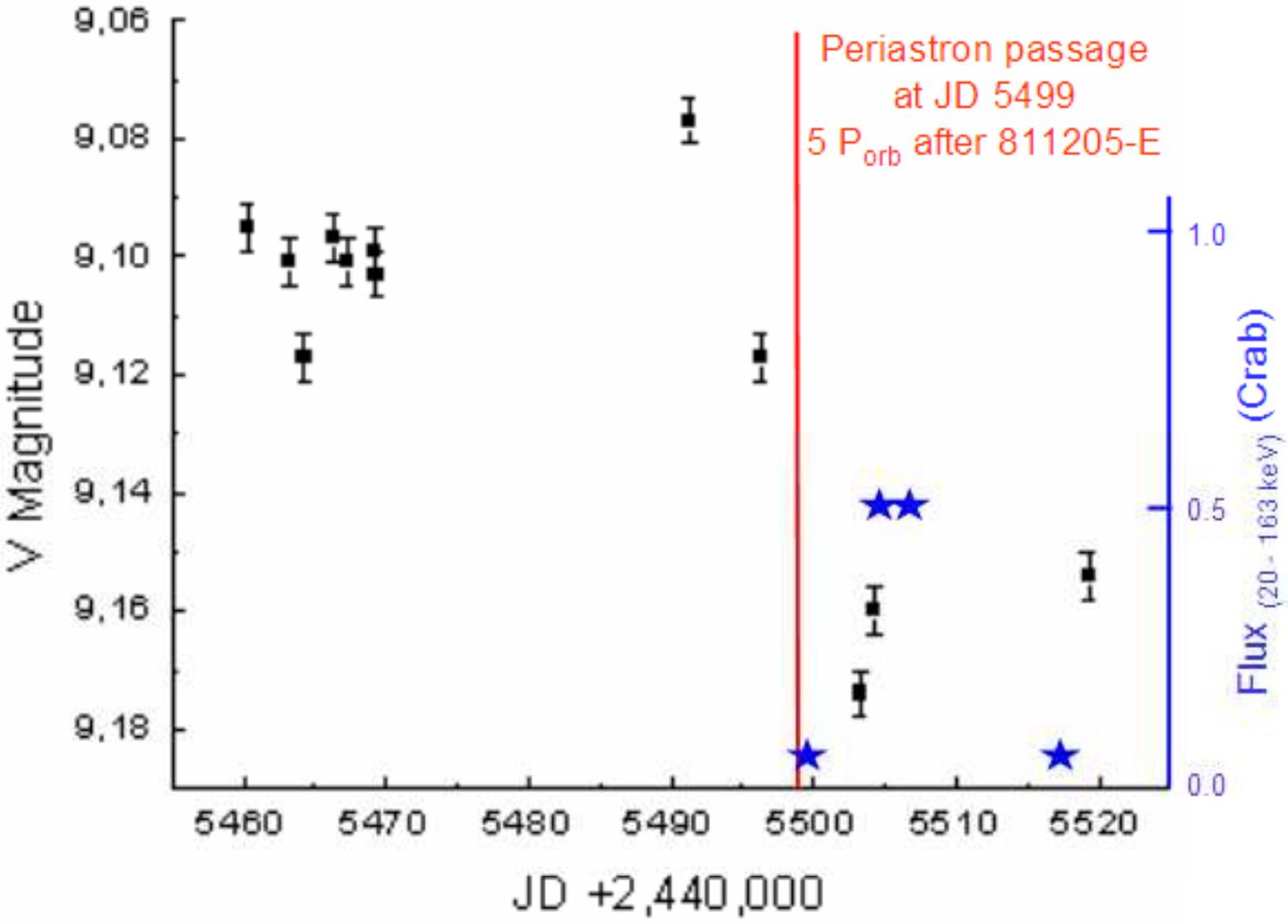}
\caption{Optical outburst V band around the 5th periastron passage after 811205-E.
Data are marked with black squares (Gnedin et al., 1988).
Red line marks the periastron passage, 5 cycle after 811205-E. X-ray data are marked with blue stars (Sembay et al., 1990, reported in GSG92).} \label{lc2}
\end{center}
\end{figure}
%%%%%%%%%%%%%%%%%%%%%%%%%%%%%%%%%%%%%%%%%%%%%%

%%%%%%%%%%%%%%%%%%% FIGURE 3 %%%%%%%%%%%%%%%%%
\begin{figure}%[!hbp]%[!ht]%[!hbp]%[h][h!]
\begin{center}
\includegraphics[width=5cm]{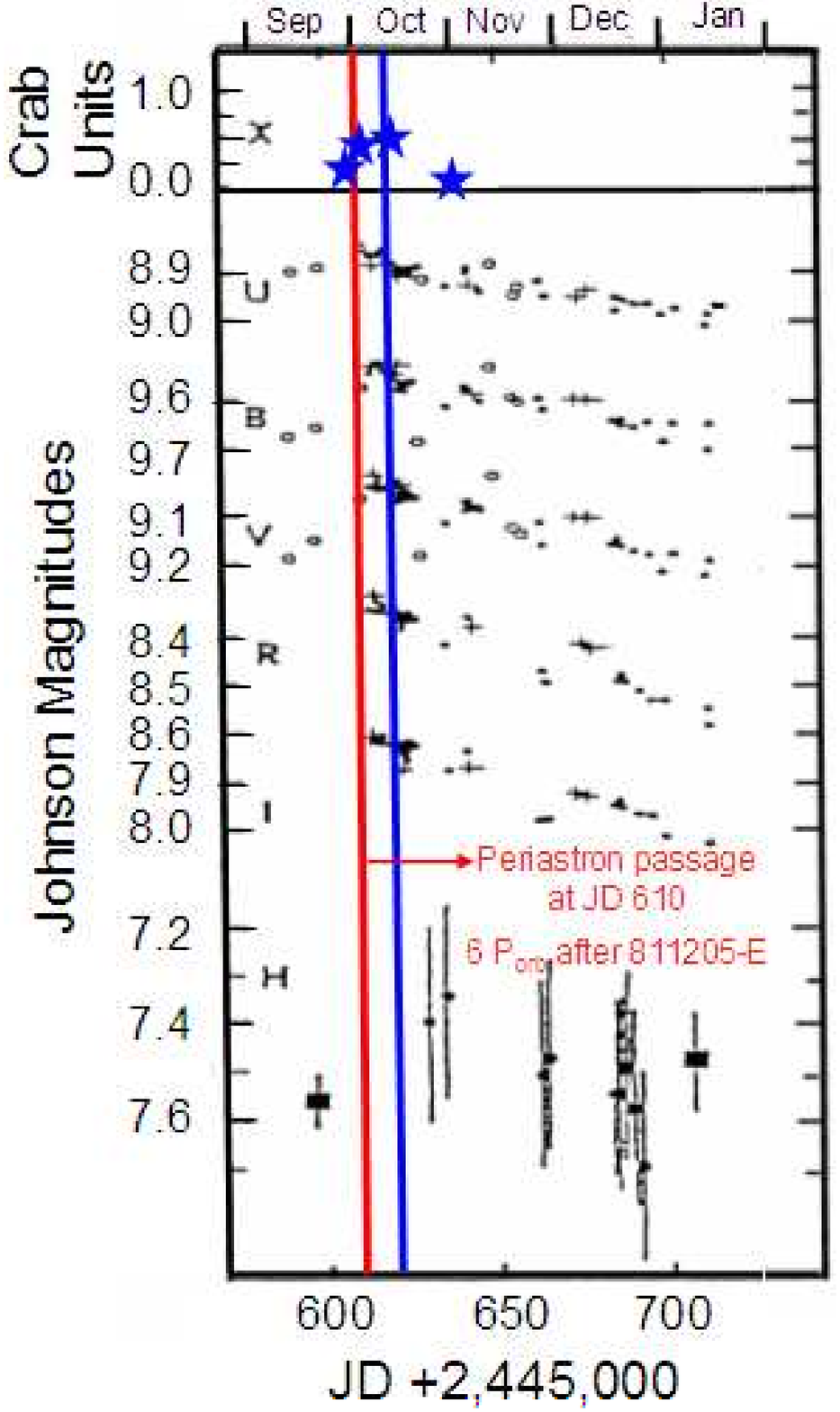}\\
\caption{Optical outburst in U,B,V,R,I, and H bands at JD 2,445,610 (Giovannelli et al., 1984) just at the 6th periastron passage after 811205-E.
Red line marks the periastron passage, 6 cycle after 811205-E; X-ray data (reported in GSG92) are marked with blue stars; the blue line marks the maximum of the X-ray outburst. } \label{lc3}
\end{center}
\end{figure}
%%%%%%%%%%%%%%%%%%%%%%%%%%%%%%%%%%%%%%%%%%%%%%

After six orbital periods (JD 2,445,609.14), we have found an optical flare in U,B,V,R,I, and H bands (Giovannelli et al., 1984) just $\sim 8$ days before the maximum of the X-ray outburst occurred in October 1--18, 1983 (GSG92). Figure 3 shows such events. The red line marks the periastron passage, and the blue line marks the maximum of the X-ray outburst. The blue stars represent the experimental points reported in GSG92, and the references therein.

%%%%%%%%%%%%%%%%%%% FIGURE 4 %%%%%%%%%%%%%%%%%
\begin{figure}%[!hbp]%[!ht]%[!hbp]%[h][h!]
\begin{center}
\includegraphics[width=8.cm]{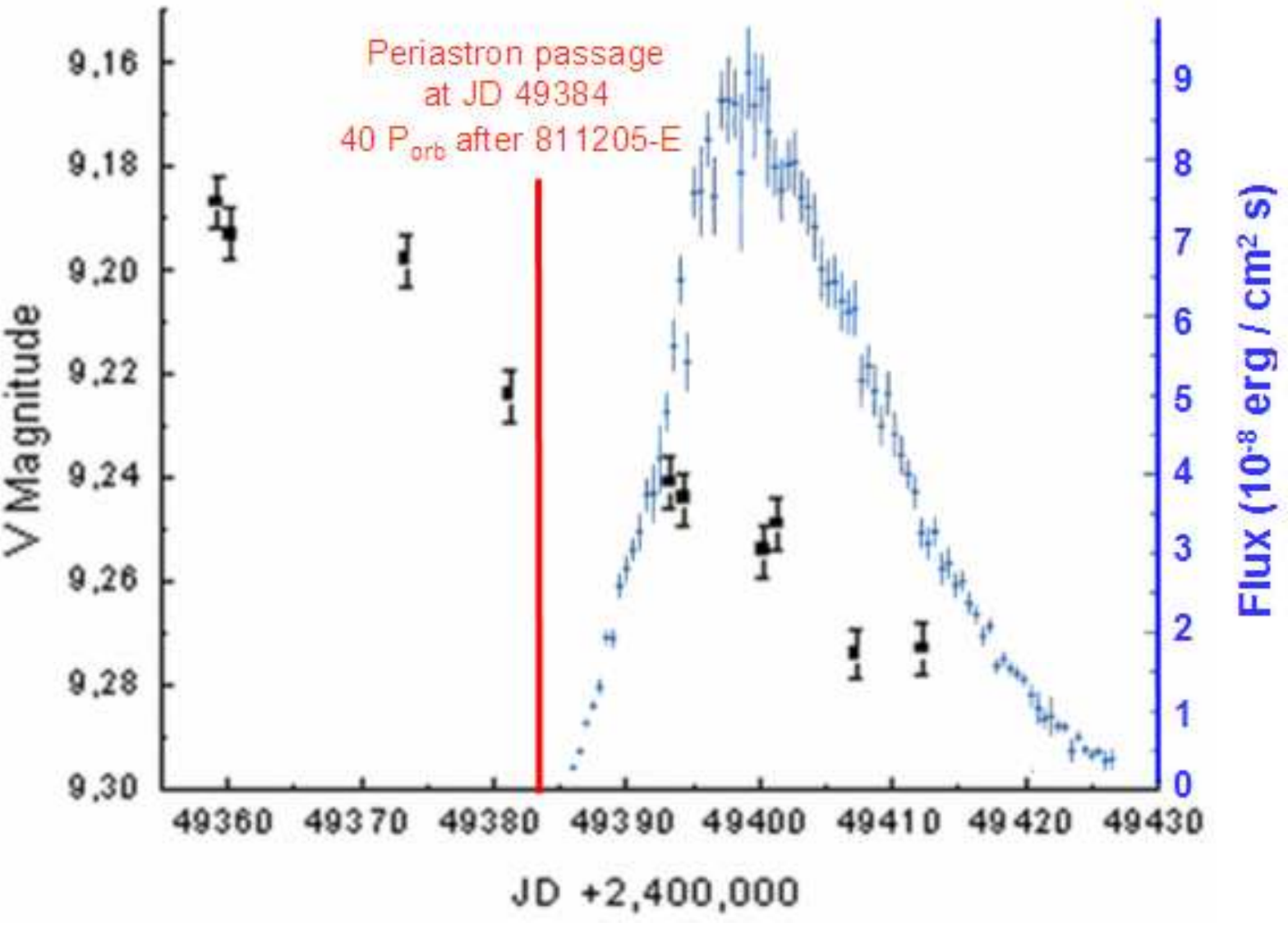}\\
\vspace{-2cm}
\caption{V magnitude around JD 2,449,384, the 40th periastron passage after 811205-E (red line). Optical data are marked with black squares (Lyuty \& Zaitseva, 2000) and the blue curve shows the subsequent X-ray outburst (Finger, Wilson \& Harmon, 1996).} \label{lc4}
\end{center}
\end{figure}
%%%%%%%%%%%%%%%%%%%%%%%%%%%%%%%%%%%%%%%%%%%%%%

%%%%%%%%%%%%%%%%%% FIGURE 5 %%%%%%%%%%%%%%%%%
\begin{figure}%[!ht] %[!hbp]%[!ht]%[!hbp]%[h][h!]
\begin{center}
\includegraphics[width=8.cm]{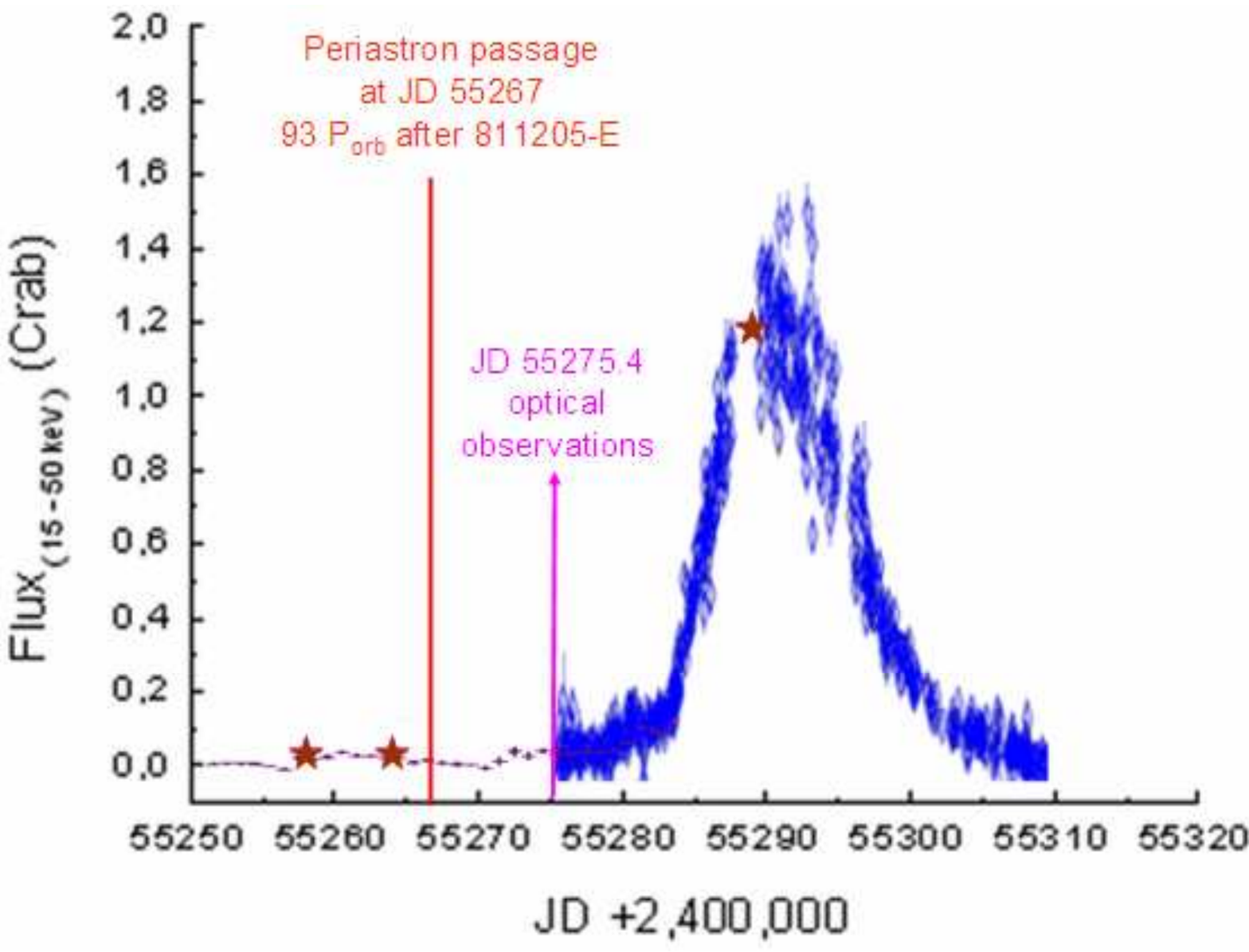}\\
\vspace{-2cm}
\caption{The 93rd periastron passage after 811205-E (red line), and the subsequent X-ray outburst detected by the Swift-BAT experiment: magenta crosses (Caballero et al., 2010c) and blue curve (Caballero et al., 2011). The magenta stars represent the experimental points of INTEGRAL (Caballero et al., 2010a,b). The epoch (JD 2,455,275.4) of spectroscopic measurements (Giovannelli, Gualandi \& Sabau-Graziati, 2010) is marked with a magenta arrow.  } \label{lc5}
\end{center}
\end{figure}
%%%%%%%%%%%%%%%%%%%%%%%%%%%%%%%%%%%%%%%%%%%%%%

%%%%%%%%%%%%%%%%%%% FIGURE 6 %%%%%%%%%%%%%%%%%
\begin{figure}%[!hbp] %[!ht] %[!hbp]%[!ht]%[!hbp]%[h][h!]
\begin{center}
\includegraphics[width=8.6cm]{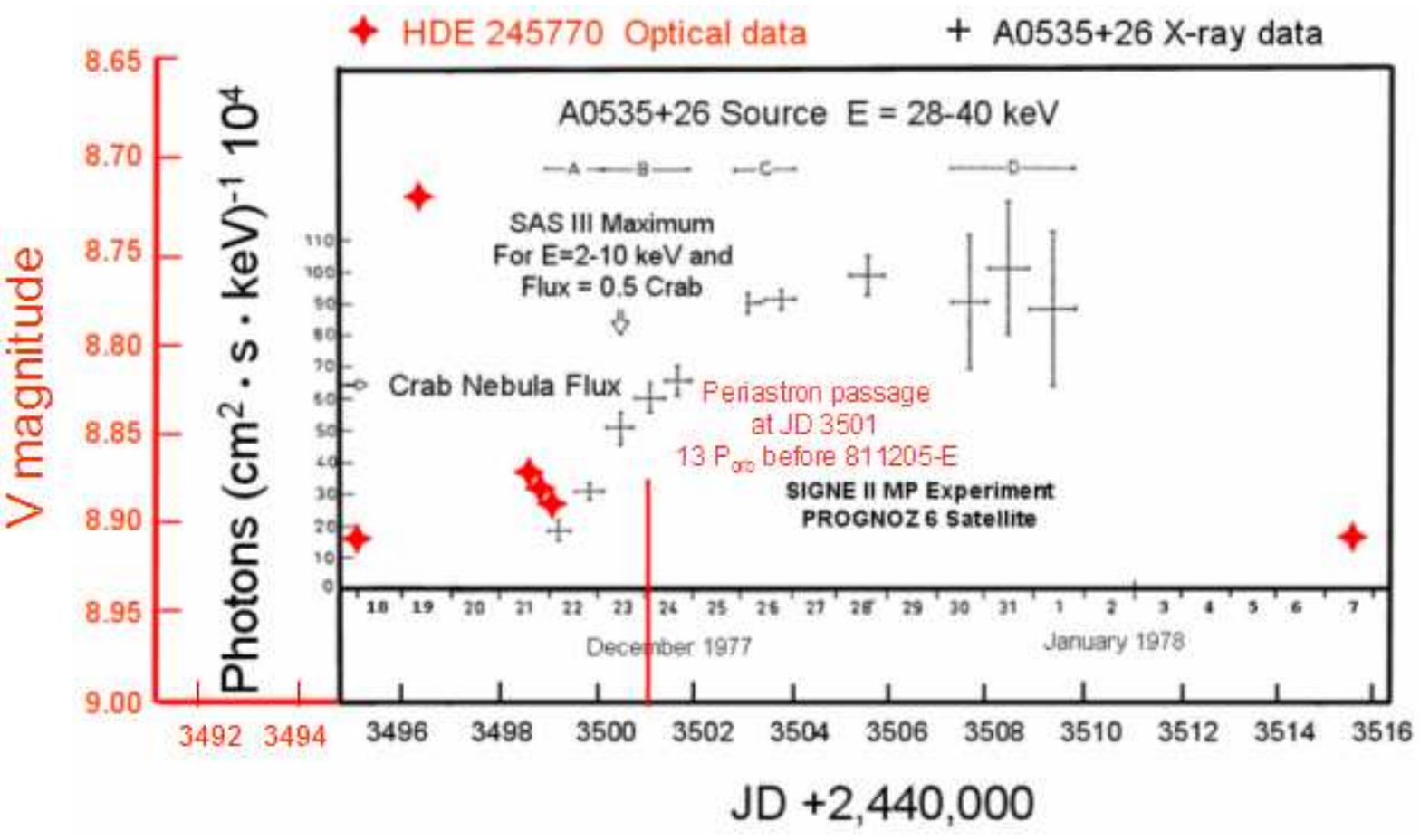}\\
\vspace{-3cm}
\caption{The optical flare of HDE 245770 (Bartolini et al., 1978; R\"{o}ssiger, 1978) (red stars) precedes the X-ray outburst of A 0535+26 (black crosses) (Violes et al., 1982). The red line marks the periastron passage at the 13th cycle before 811205-E. } \label{lc6}
\end{center}
\end{figure}
%%%%%%%%%%%%%%%%%%%%%%%%%%%%%%%%%%%%%%%%%%%%%%

%%%%%%%%%%%%%%%%%%% FIGURE 7 %%%%%%%%%%%%%%%%%
\begin{figure}%[!ht]%[!hbp]%[!ht]%[!hbp]%[!ht] %[!hbp]%[!ht]%[!hbp]%[h][h!]
\begin{center}
\vspace{-2cm}
\includegraphics[width=6cm]{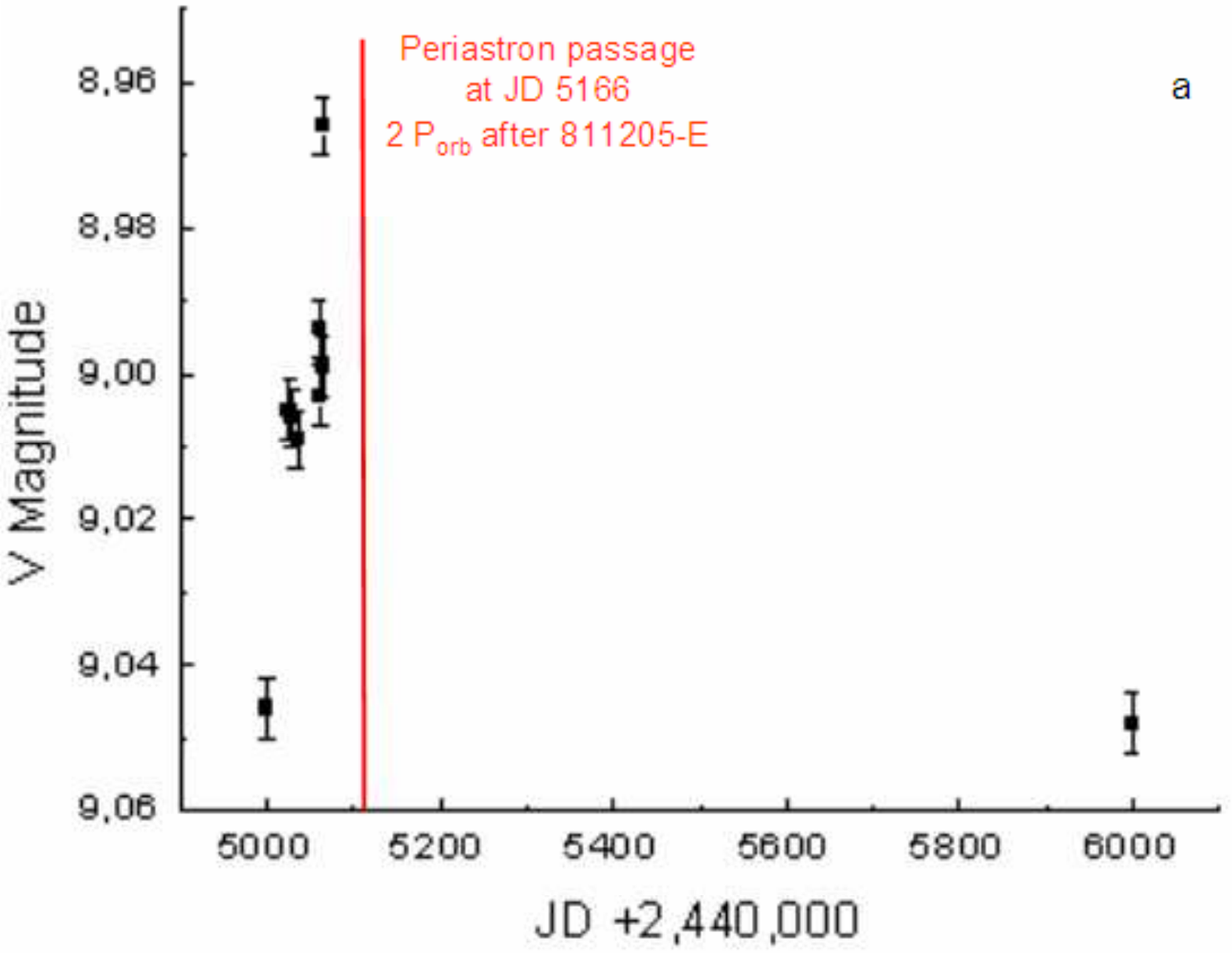}
\includegraphics[width=6cm]{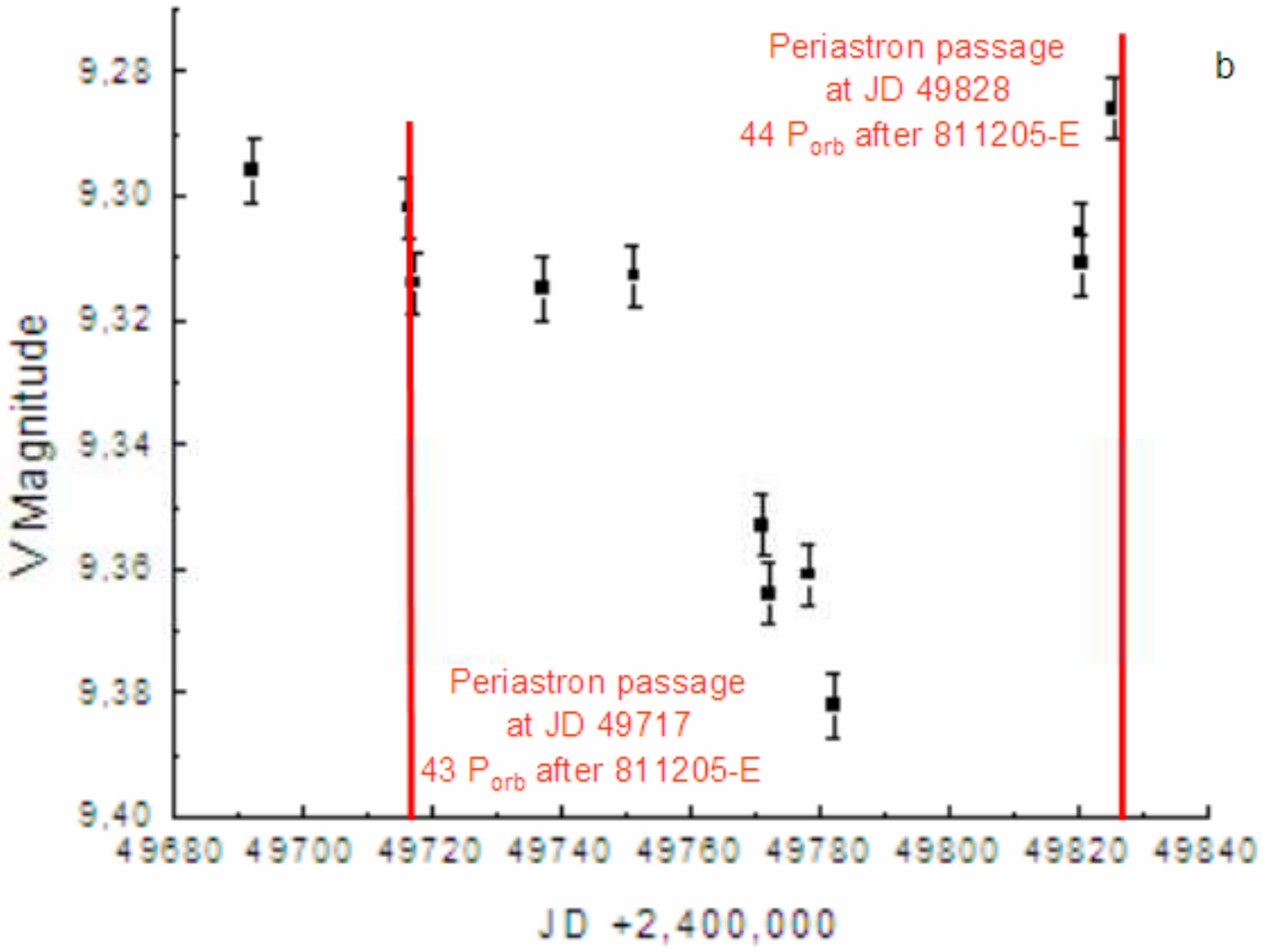}
\vspace{-2cm}
\includegraphics[width=6cm]{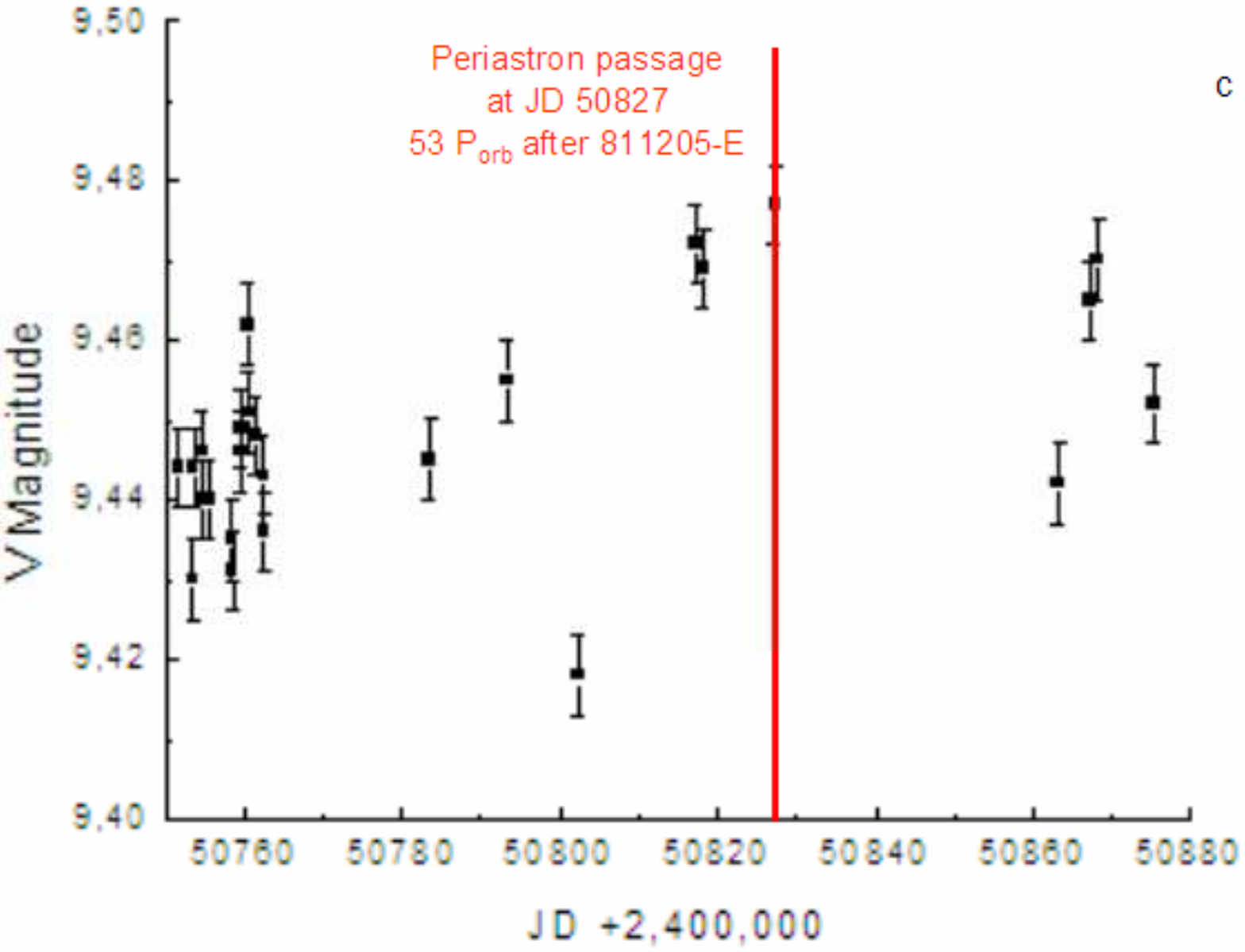}
\includegraphics[width=6cm]{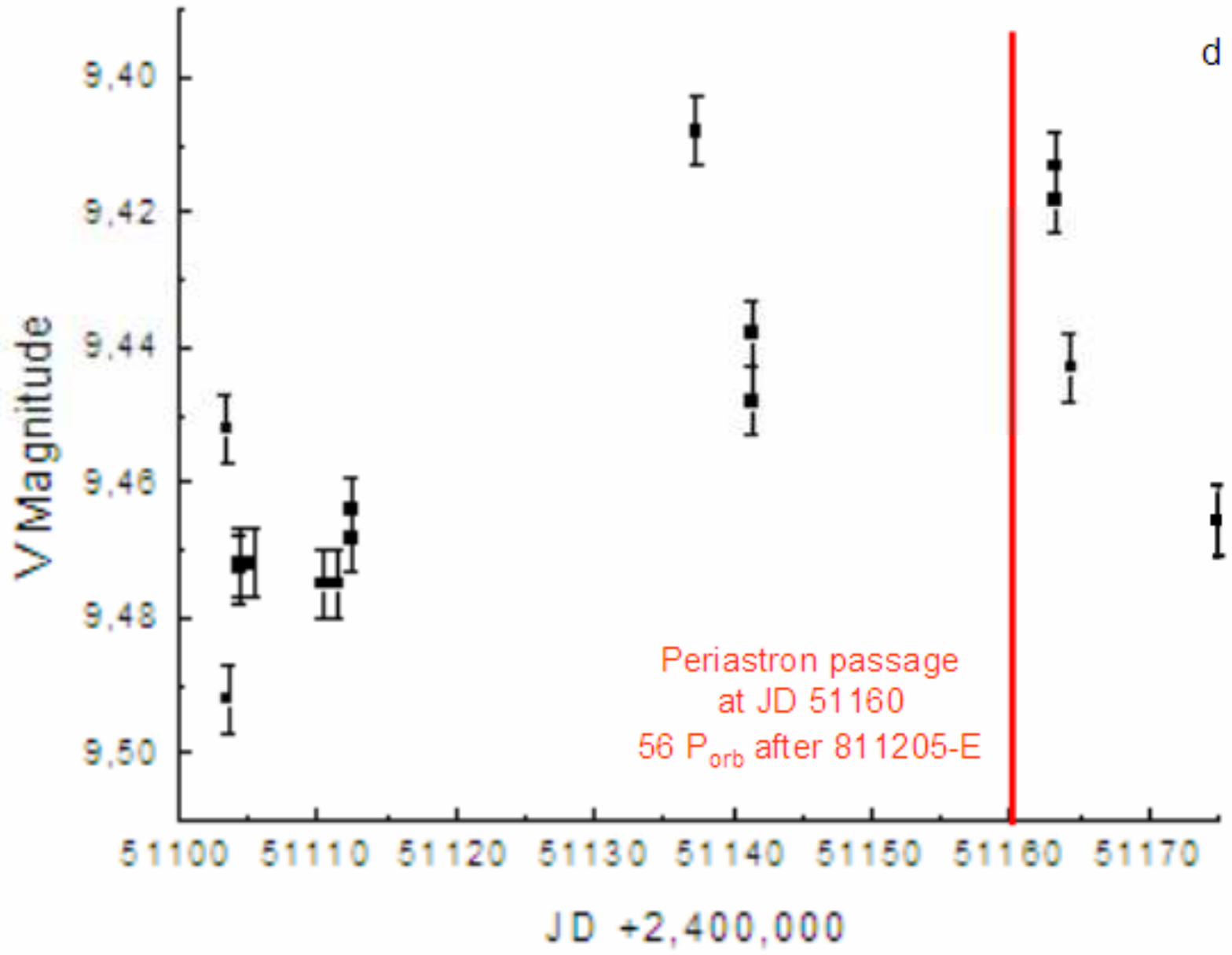}
%\vspace{-2cm}
\caption{V magnitudes of HDE 245770 at various epochs (black squares). Red lines mark the periastron passages at correspondent cycles after 811205-E.
Optical data are taken from Lyuty \& Zaitseva (2000), but the data reported in the upper left panel, which are taken from Gnedin et al. (1988). } \label{lc7}
\end{center}
\end{figure}
%%%%%%%%%%%%%%%%%%%%%%%%%%%%%%%%%%%%%%%%%%%%%%

Figure 4 shows optical data (Lyuty \& Zaitseva, 2000) and X-ray data (Finger, Wilson \& Harmon, 1996) at the 40th cycle (JD 2,449378.240) after 811205-E.
The optical luminosity at the periastron passage is just starting the decay from a relative maximum, whilst the X-ray outburst starts $\sim 1$ day later and reaches about half of its maximum intensity $\sim 8$ days after the periastron passage.  This fact will be commented later in this section after the description of Fig. 9.

Figure 5 shows the strong 2010 March--April X-ray outburst of A0535+26 (Caballero et al., 2010a,b,c; Caballero et al., 2011). This strong outburst was predicted by Giovannelli, Gualandi \& Sabau-Graziati (2010) on the basis of a strong activity of HDE 245770 manifested in the Balmer lines. The 93rd cycle after 811205-E, marked with a red line, precedes $\sim 8$ days the beginning of the X-ray outburst detected by the  SWIFT-BAT experiment.

Analyzing the first  five figures it appears evident that X-ray outbursts follow the periastron passage of $\sim 8$ days. In order to enforce such conclusion we show what happened at the epoch of December 1977 -- January 1978 X-ray outburst of A 0535+26 (Violes et al., 1982).  Just a few days before the beginning of the X-ray outburst, expected by Giovannelli's group, as described in Giovannelli \& Sabau-Graziati (2011), Bartolini et al. (1978) detected a decay from an optical flare of HDE 245770, which rendered possible its final association with the X-ray pulsar A 0535+26. Independent of Bartolini et al. (1978) measurements, R\"{o}ssiger (1978) detected the maximum of the optical flare at JD +2,443,496.370.
Figure 6 shows the optical flare (red stars) constructed by using Bartolini et al. (1978) and R\"{o}ssiger (1978) data, and the subsequent X-ray outburst (black crosses) (Violes et al., 1982). There is full consistency about the fact that the optical flare precedes the beginning of the X-ray outburst of several days, in this case $\sim 3$. The periastron passage, at 13 cycles before 811205-E is marked with a red line. This passage precedes the maximum of the X-ray outburst in the range 28--40 keV of  $\sim 5$ days.

 Figure 7 shows a mosaic of plots in which are reported the V magnitudes of HDE 245770 (Flavia' star) in various epochs in correspondence of the cycles 2, 43, 44, 53, and 56 after 811205-E. In each plot a red line marks the passage at the periastron corresponding to the proper cycle after 811205-E. The source of the optical data used are from Lyuty \& Zaitseva (2000), except the data reported in Fig. 7a, which are taken from Gnedin et al. (1988). Unfortunately, X-ray data are not available for those epochs. However, at the periastron passage the optical luminosity is at a relative maximum or very close.
Looking at the Fig. 7a it is possible to note that the optical flare, whose maximum is reached at JD 2,445062.267, has a magnitude of order 0.1. A similar optical outburst detected in December 1977 (see Fig. 6), with a magnitude of order 0.17, whose maximum was reached at JD 2,443,496.370, is separated by the former just 14.11 P$_{\rm orb}$.

Finally, we can show several events detected in X-ray energy range, without correspondent optical data.

 Three cycles after 811205-E, the epoch of X-ray data from A0535+26 -- in the range 1--22 keV from the Hakucho satellite (Nagase et al., 1984) and reported in the Table 1 of GSG92 -- follows the periastron passage of $\sim 9$ days. The information available is that the X-ray flux was of $\sim 0.2$ Crab on November 12--20, 1982 (JD 2,445,286--294) (see  Fig. 8). Upper and lower panels of Fig. 8 report R and I data together with X-ray data, respectively; red line marks the periastron passage. It is remarkable to note that the detected X-ray emission (normal outburst, following GSG92 classification) occurs $\sim 9$ days after the periastron passage. Unfortunately the measurements in R an I bands of HDE 245770, made by Gnedin et al. (1988), are not available at the periastron.

%%%%%%%%%%%%%%%%%%% FIGURE 8 %%%%%%%%%%%%%%%%%
\begin{figure} %[!hbp]%[!ht]%[!hbp]%[h][h!]
\begin{center}
\includegraphics[width=6cm]{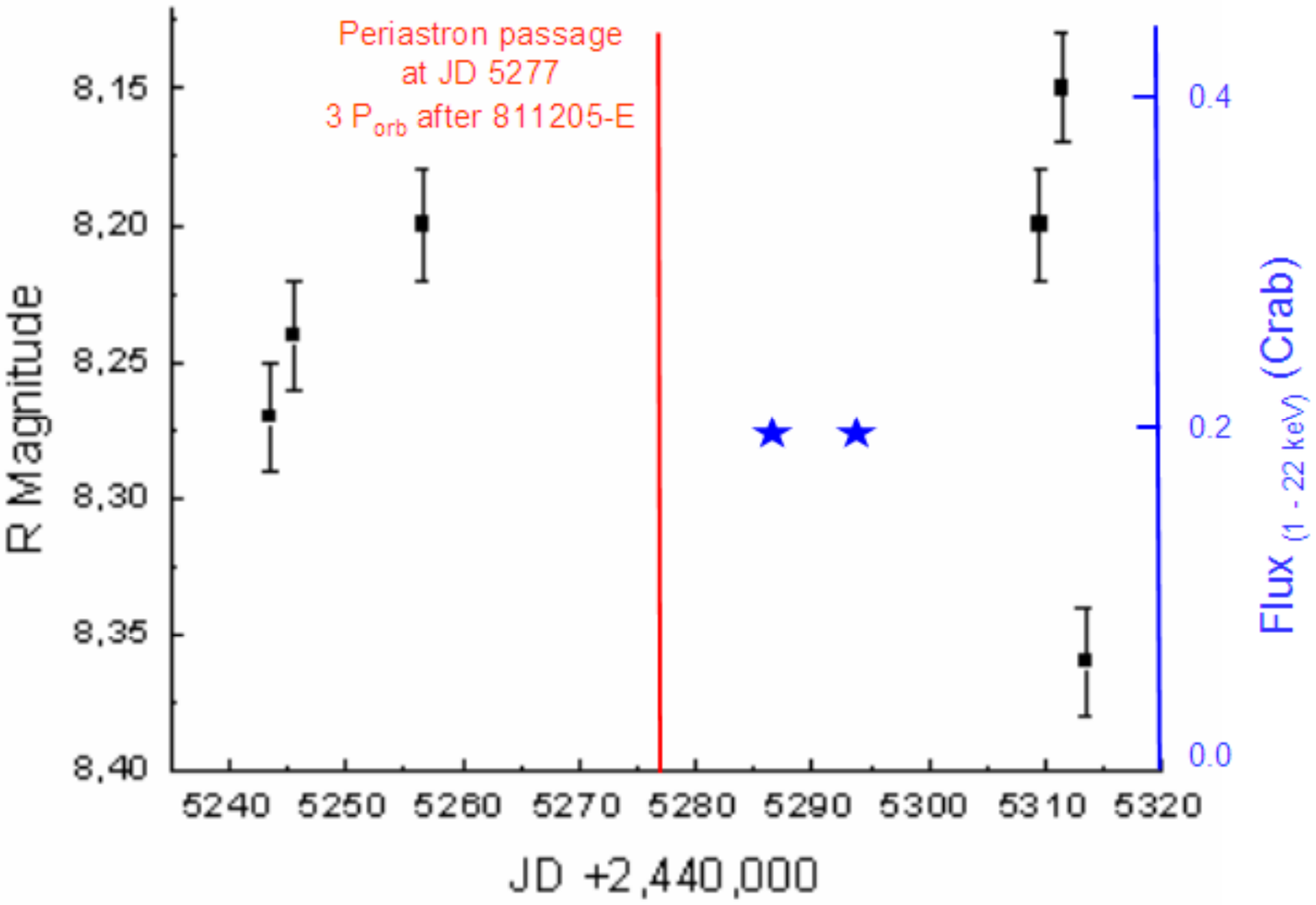}
\vspace{-2cm}
\includegraphics[width=6cm]{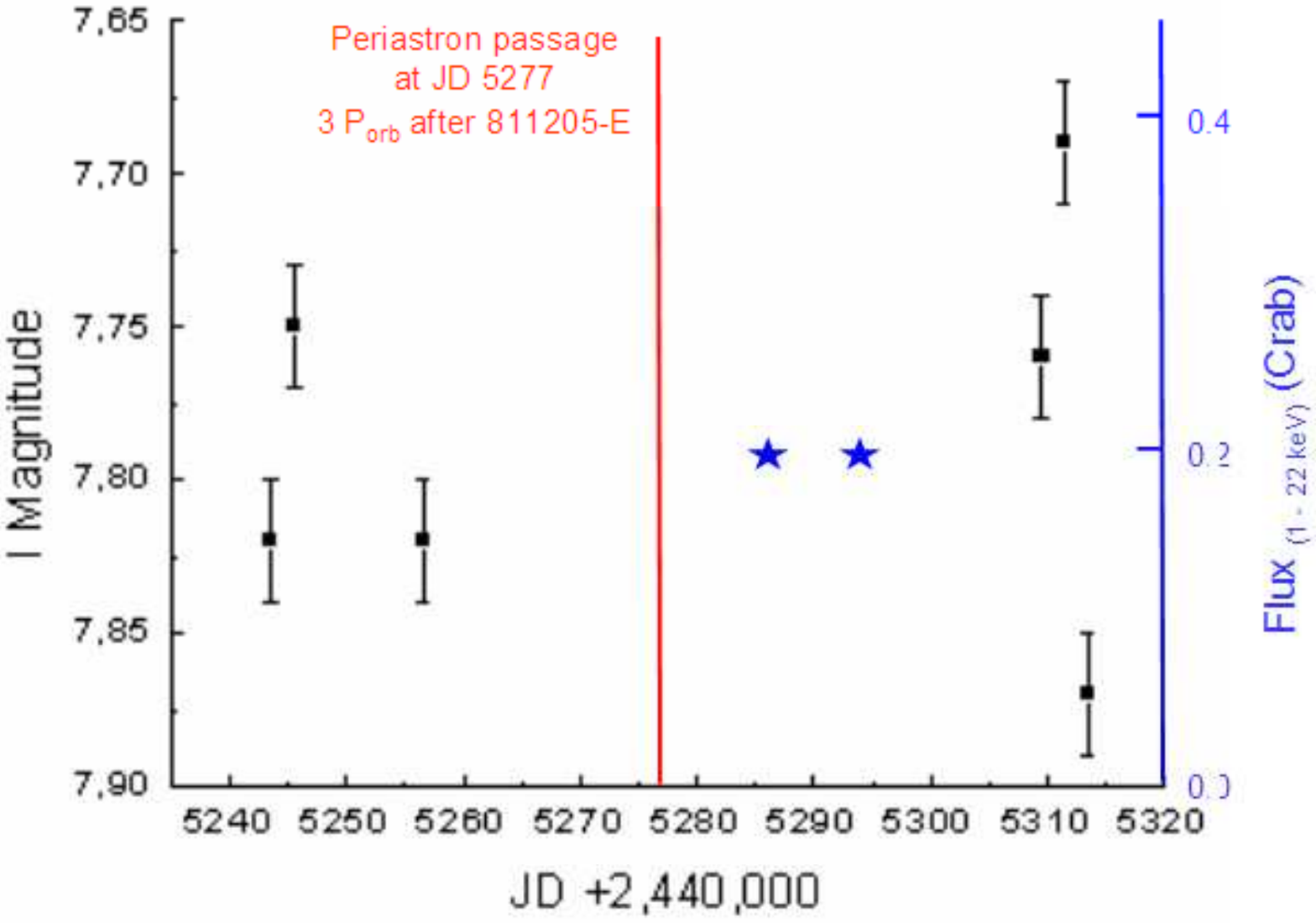}
\caption{Optical probable outburst in R and I bands (left and right panel, respectively) around the 3th periastron passage after 811205-E.
Optical data are marked with black squares (Gnedin et al., 1988).
Red line marks the periastron passage, 3 cycle after 811205-E. X-ray data are marked with blue stars (Nagase et al., 1984, reported in GSG92).} \label{lc8}
\end{center}
\end{figure}
%%%%%%%%%%%%%%%%%%%%%%%%%%%%%%%%%%%%%%%%%%%%%%

 Close to the 7th cycle after 811205-E (19th January 1984 $\rightarrow$ JD 2,445,720), X-ray data from ASTRON satellite in the range 2 -- 25 keV (Giovannelli et al., 1984), reported in GSG92, are available.
Figure 9 reports V data (upper panel) and J data (lower panel), marked in black, unfortunately not at the periastron,  together with the two X-ray points available (JD 2,445,732 and 735), marked with blue stars.
It is remarkable to note that the first X-ray point of the decay of the outburst is placed 11 days after the periastron passage. Although the optical data are not available at the periastron passage, from the upper panel of Fig. 7 we can argue that the V luminosity is decaying from a relative maximum value, when the X-ray outburst appears.

This is the third event, together with those reported in Fig. 2 and Fig. 4 where X-ray outbursts develop during the decay of optical luminosity. This is also the case of the December 77--January 78 strong X-ray outburst that starts during the decay of the optical luminosity (see Fig. 6). This experimental evidence is corroborated by the finding of Yan, Li \& Liu (2012) who found that each giant X-ray outburst occurred in a fading phase of the optical brightness. Moreover, the anti-correlation between the optical brightness and H$_\alpha$ intensity during their 2009 observations indicates a mass ejection event had taken place before the July--August 2009 giant X-ray outburst, reported in Fig. 11b and classified as casual or strong X-ray outburst by GSG92.

%%%%%%%%%%%%%%%%%%% FIGURE 9 %%%%%%%%%%%%%%%%%
\begin{figure} %[!hbp]%[!ht]%[!hbp]%[h][h!]
\begin{center}
\includegraphics[width=6cm]{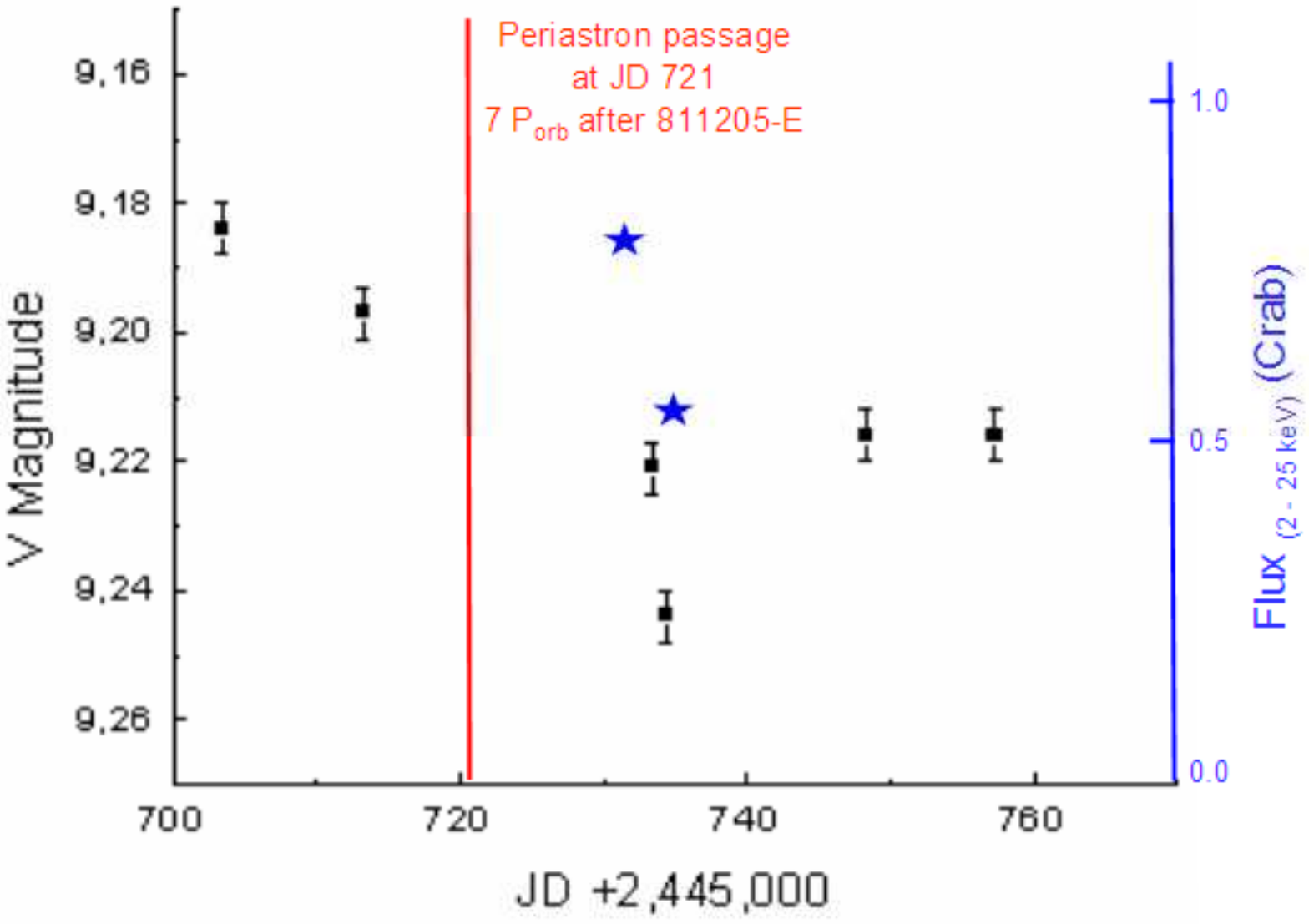}
\vspace{-2cm}
\includegraphics[width=6cm]{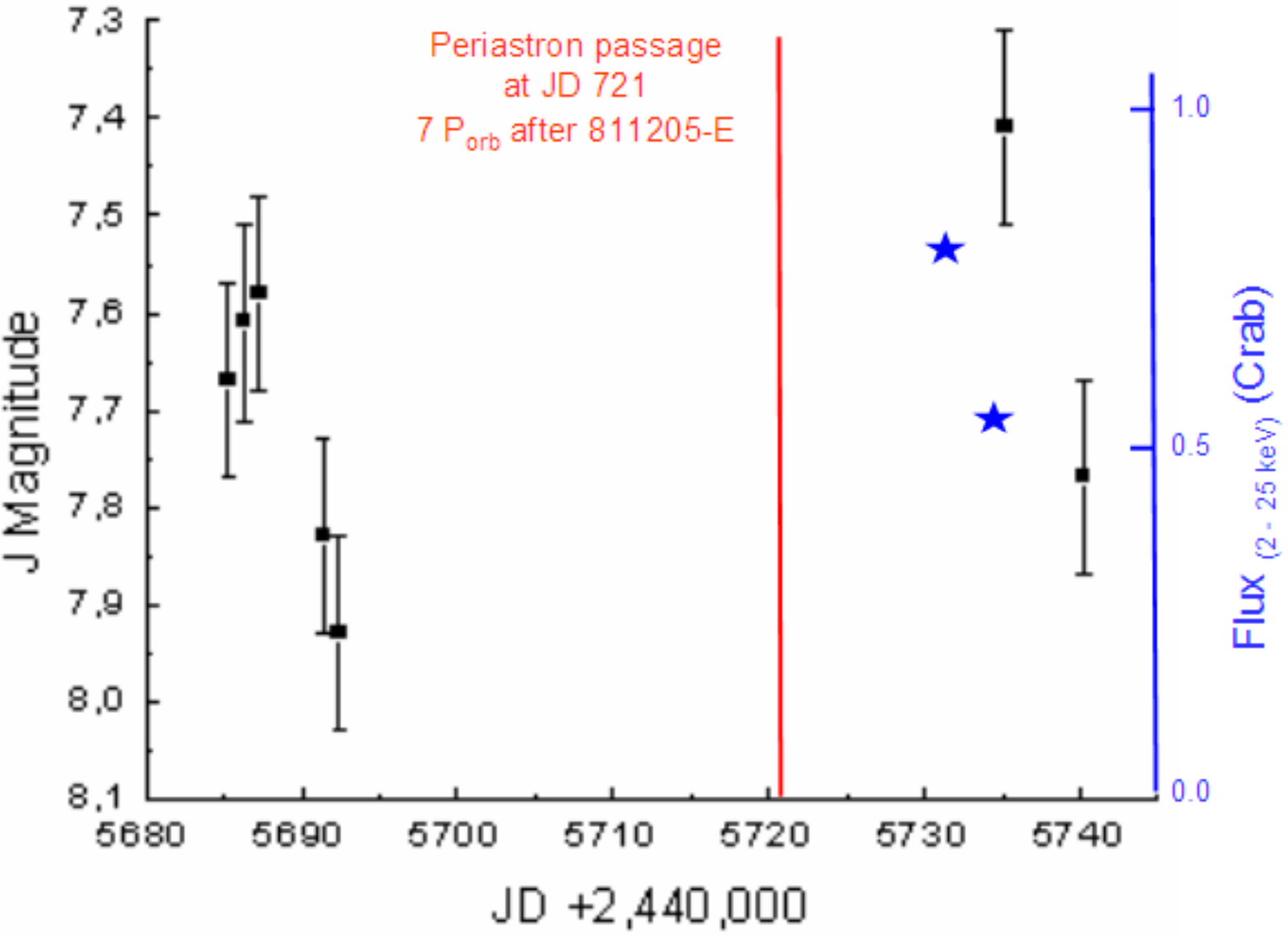}
\caption{Optical data in V and J bands (black squares) (left and right panels, respectively) (Gnedin et al., 1988) together with X-ray data (blue stars) (Giovannelli et al., 1984, reported in GSG92), around the 7th periastron passage after 811205-E.
Red line marks the periastron passage, 7 cycle after 811205-E. } \label{lc9}
\end{center}
\end{figure}
%%%%%%%%%%%%%%%%%%%%%%%%%%%%%%%%%%%%%%%%%%%%%%

Figure 10 shows the 17th September, 1980 periastron passage which is 4 cycles before 811205-E. Red line marks the periastron passage and blue rectangle the epoch of the September--October X-ray outburst, detected by several experiments whose results are reported in GSG92, and references therein. The few U, B, and V experimental points are reported in the paper by Giovannelli et al. (1985).  Unfortunately no optical data are available at the periastron.
It is impressive to note that the X-ray outburst follows the periastron passage exactly of 8 days. The start of the X-ray outburst is just 4 orbital periods before the 811213-E, the X-ray short flare that followed the 811205-E. We can say that we are in the presence of a spectacular clock!

%%%%%%%%%%%%%%%%%%% FIGURE 10 %%%%%%%%%%%%%%%%%
\begin{figure}[!ht] %[!hbp]%[!ht]%[!hbp]%[h][h!]
\begin{center}
\includegraphics[width=5.5cm]{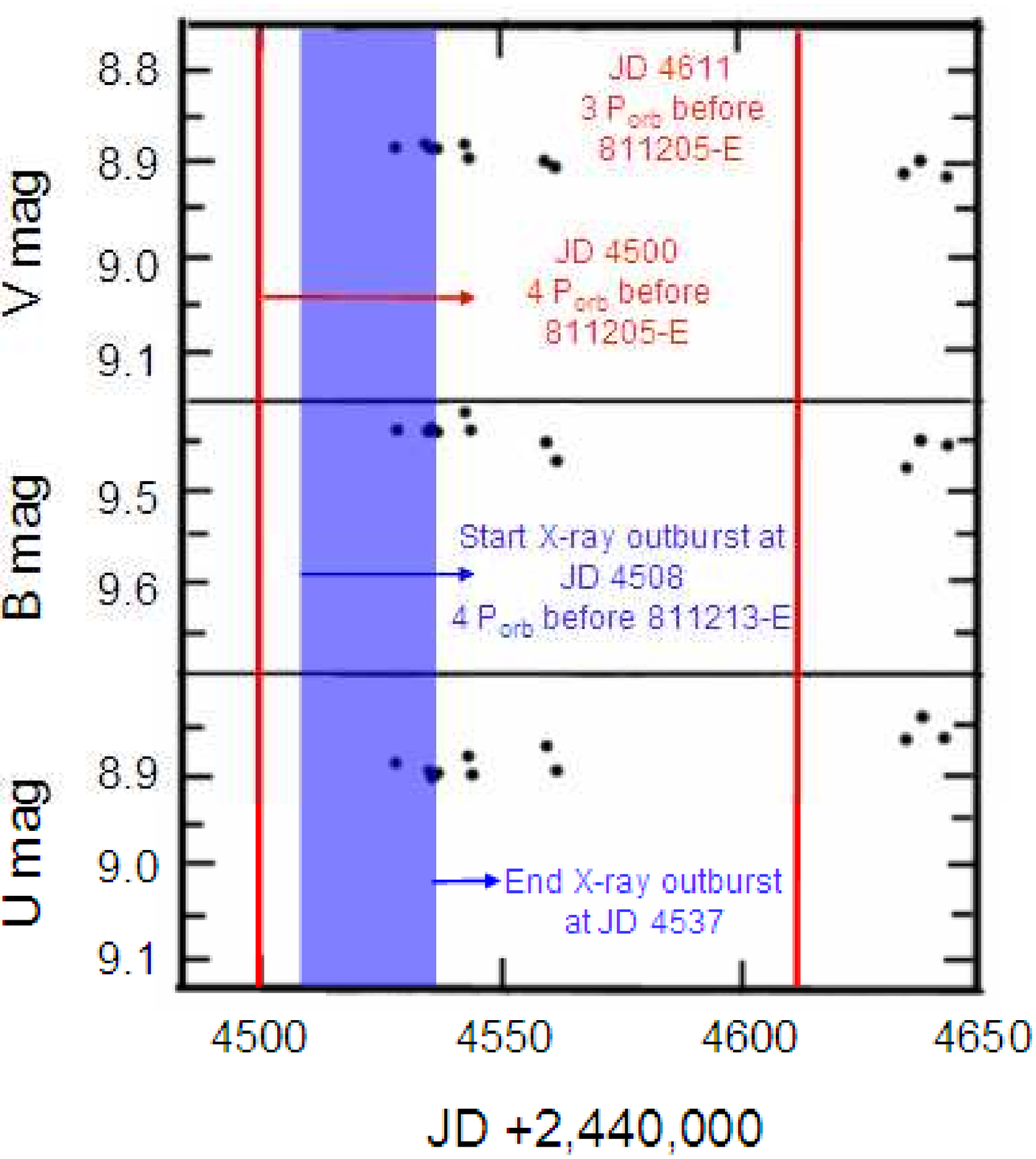}\\
\caption{The Periastron passage at the 4th cycle before 811205-E (JD 4500) (red line) precedes of $\sim 8$ days the X-ray outburst of A 0535+26 that starts on JD 4508 ending at JD 4537 (blue rectangle) (GSG92, and the references therein). Optical U, B, and V data (black points) are from Giovannelli et al. (1985).} \label{lc10}
\end{center}
\end{figure}
%%%%%%%%%%%%%%%%%%%%%%%%%%%%%%%%%%%%%%%%%%%%%%

Figure 11 shows a mosaic of plots in which it is possible to look at various X-ray outbursts starting $\sim 8$ days after the correspondent periastron passage.  Panel `a' shows the 2005 August--September X-ray outburst (Caballero et al., 2007). The periastron passage at the 78th cycle after 811205-E is $\sim 8$ days before the X-ray outburst.  Panel `b' shows the 2009 July--August X-ray outburst (Caballero et al., 2010d). The periastron passage at the 91st cycle after 811205-E is $\sim 3-12$ days before the double peaked X-ray outburst. Panel `c' shows the 2009 November--December X-ray outburst (Caballero et al., 2011). The periastron passage at the 92nd cycle after 811205-E is $\sim 10$ days before the rise of the X-ray outburst. Panel `d' shows the 2010 June--August X-ray outburst (Camero-Arranz et al., 2012). The periastron passage at the 94th cycle after 811205 (JD 55364.464) is coincident with the first peak of the double peaked X-ray outburst and $\sim 30$ days befo
 re the second peak. This outburst could be an exception, that deserves a deep analysis.

%%%%%%%%%%%%%%%%%%% FIGURE 11 %%%%%%%%%%%%%%%%%
\begin{figure}[!ht]%[!hbp]%[!ht] %[!hbp]%[!ht]%[!hbp]%[h][h!]
\begin{center}
\vspace{-2cm}
\includegraphics[width=6cm]{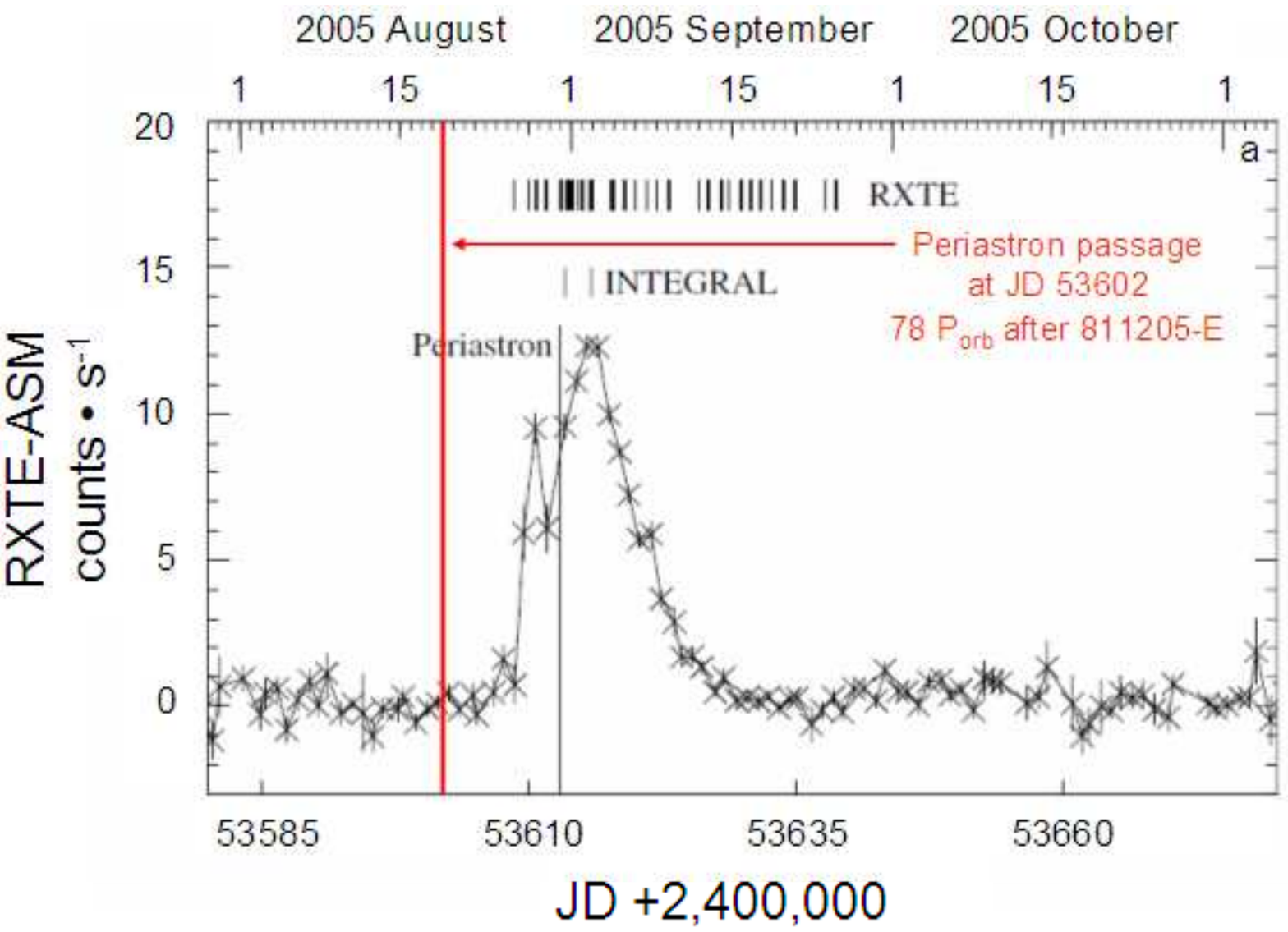}
%\vspace{-4.6cm}
\includegraphics[width=6cm]{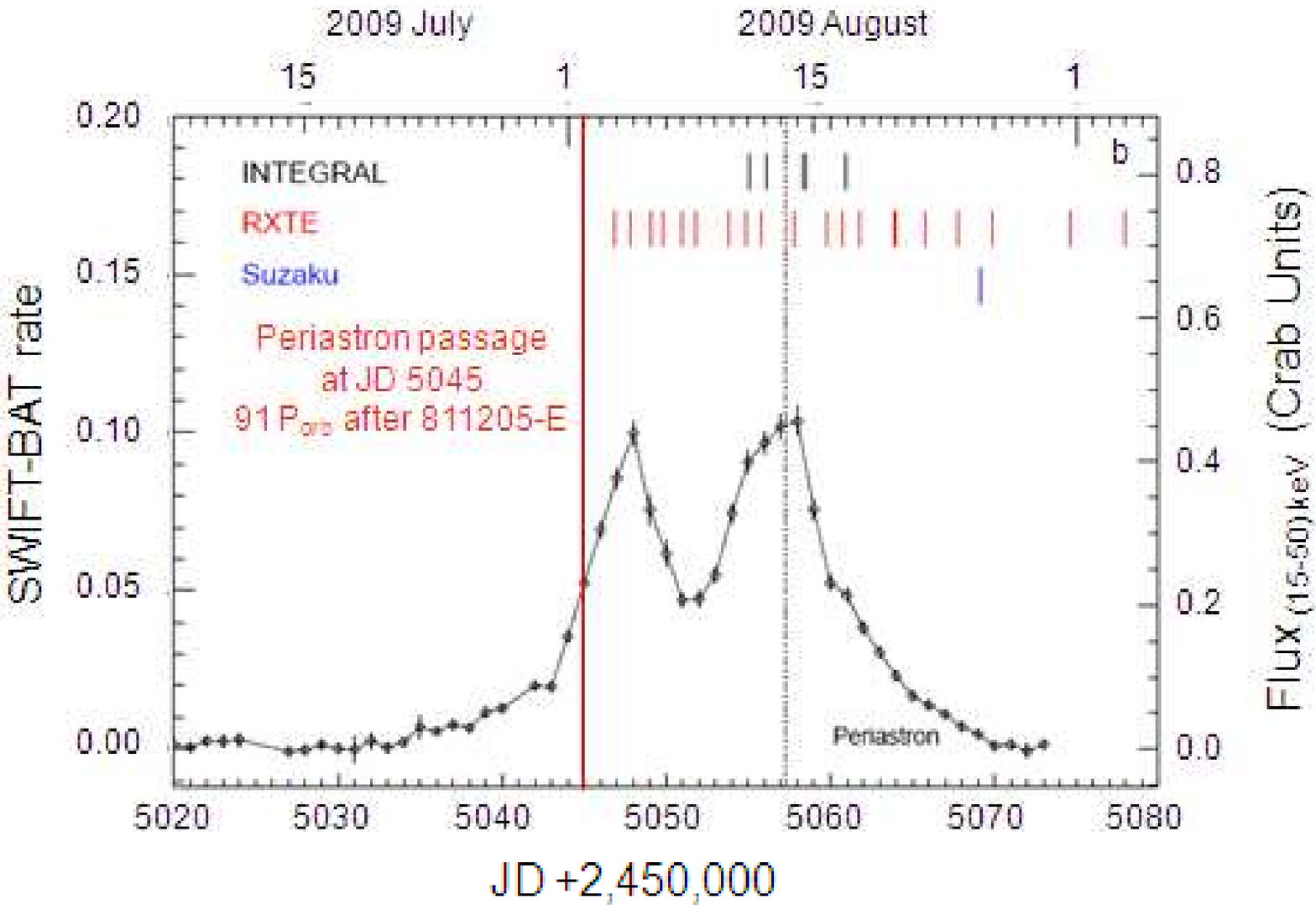}\\
\vspace{-4.6cm}
\includegraphics[width=6cm]{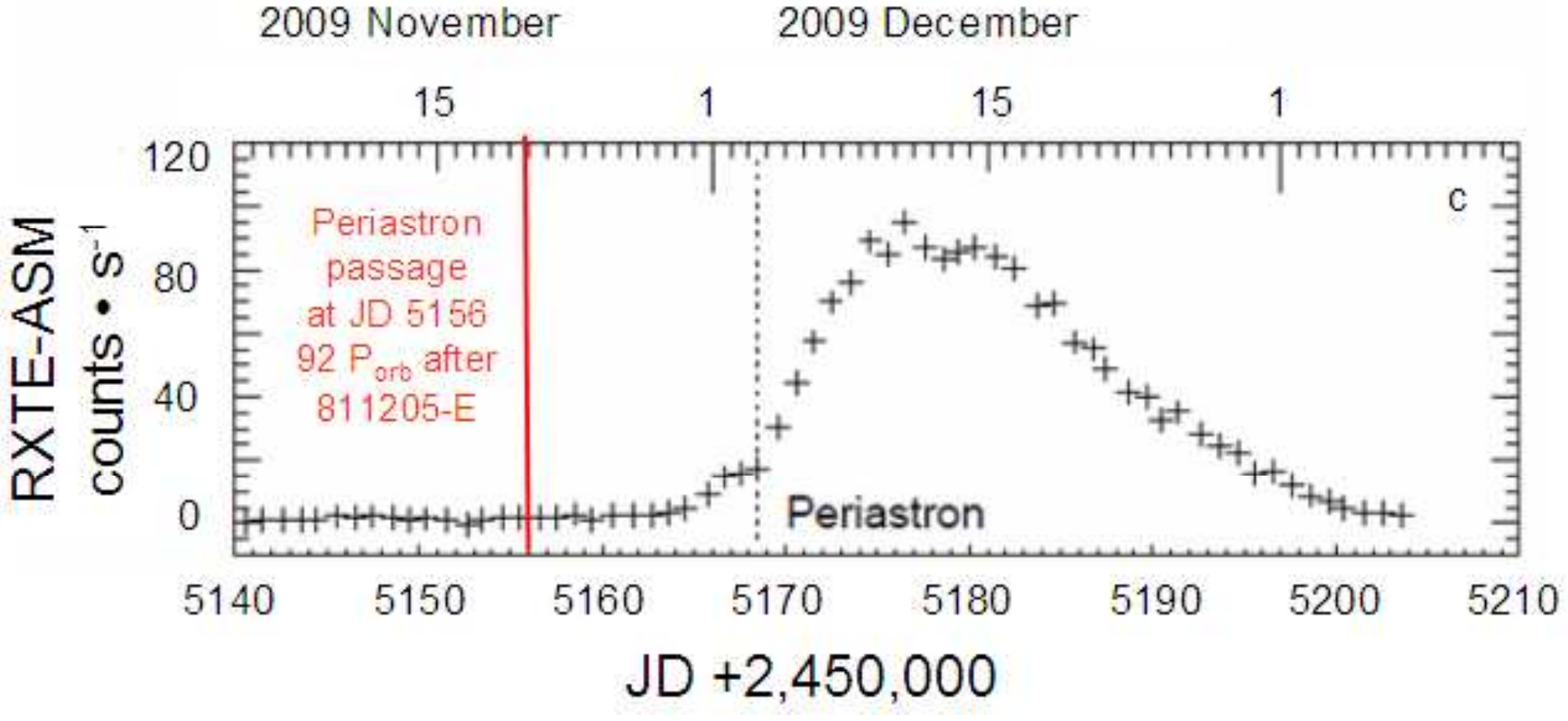}
%\vspace{-4.6cm}
\includegraphics[width=6cm]{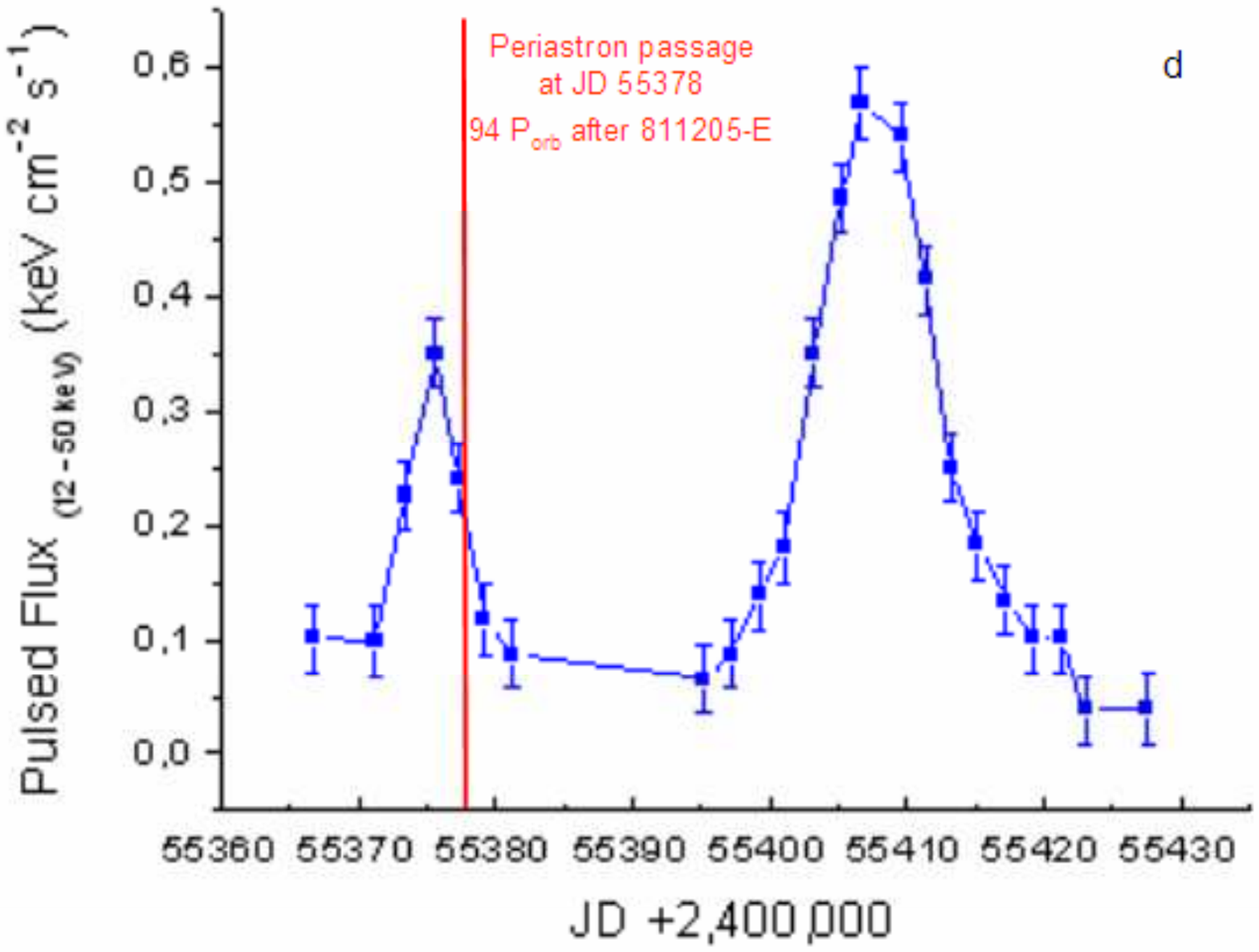}\\
%\vspace{2.5cm}
\caption{Panel (a) up-left: 2005 August--September X-ray outburst (after Caballero et al., 2007);
panel (b) up-right: 2009 July--August X-ray outburst (after Caballero et al., 2010; panel (c) down-left: 2009 November--December X-ray outburst (after Caballero et al., 2011); panel (d) down-right: 2010 June--August X-ray outburst (after Camero-Arranz et al., 2012). Red lines mark the relative periastron passages at 78th, 91st, 92nd, and 94th cycles after 811205-E, respectively.} \label{lc11}
\end{center}
\end{figure}
%%%%%%%%%%%%%%%%%%%%%%%%%%%%%%%%%%%%%%%%%%%%%%

%%%%%%%%%%%%%%%%%%% FIGURE 12 %%%%%%%%%%%%%%%%%
\begin{figure*}[ht]%1[!hbp]%[!ht] %[!hbp]%[!ht]%[!hbp]%[h][h!]
\vspace{-3cm}
\begin{center}
\includegraphics[width=12.6cm]{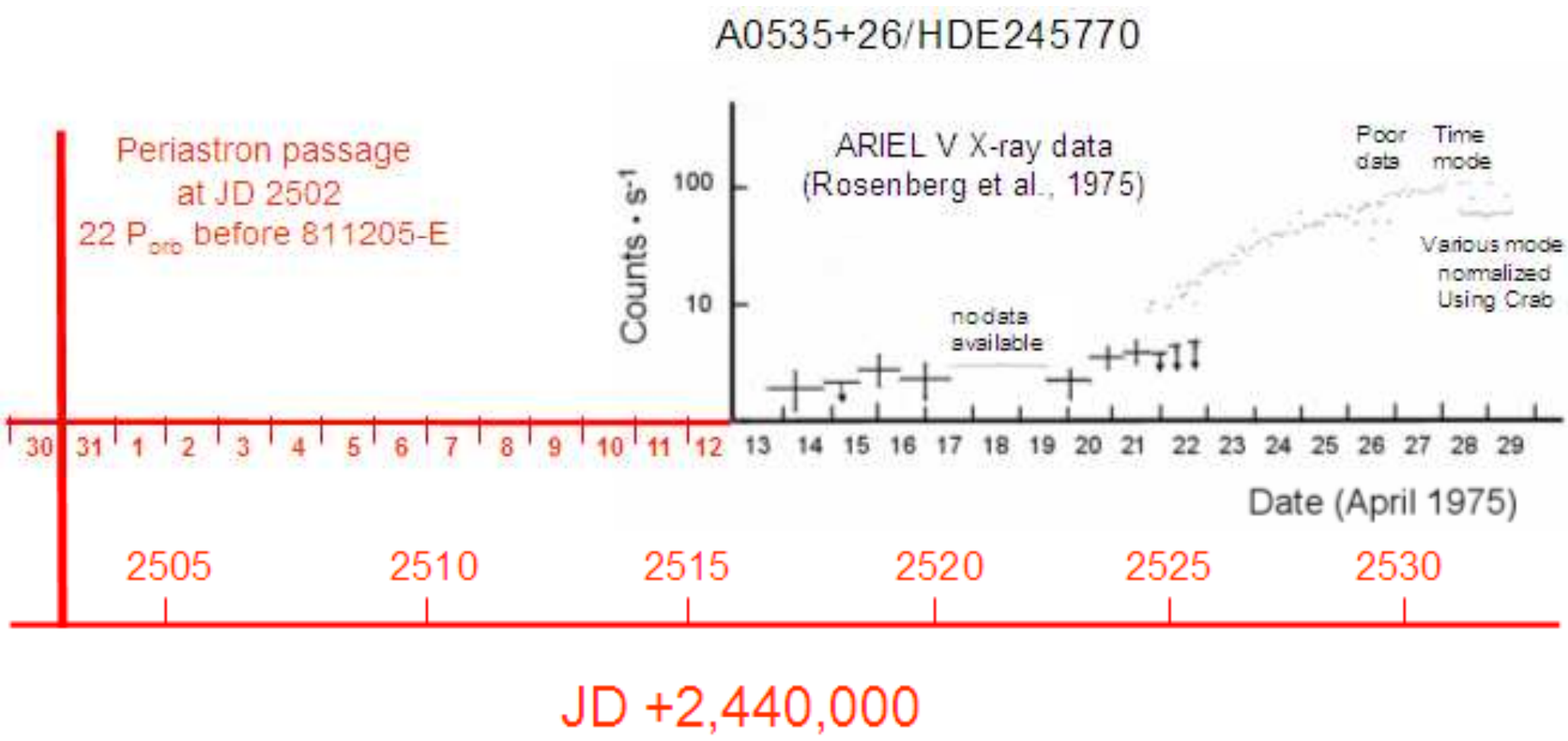}\\
\vspace{-5cm}
\caption{The Periastron passage at the 22nd cycle before 811205-E (JD 2502) (red line) precedes of $\sim 14$ days the X-ray outburst of A 0535+26 that starts approximatively on JD 2516 (after Rosenberg et al., 1975).} \label{lc12}
\end{center}
\end{figure*}
%%%%%%%%%%%%%%%%%%%%%%%%%%%%%%%%%%%%%%%%%%%%%%

In order to have a final strong proof about such a delay, we have used the data of the first detection of A0535+26 (Rosenberg et al., 1975). Figure 12 shows the first X-ray emission detected from A0535+26 (black crosses and points) and the periastron passage at the 22nd cycle before 811205-E (red line). The X-ray detection of the pulsar starts  $\sim 14$ days after the periastron passage. No X-ray data are available before.
It is very impressive how the clock is working!

 Thus, in conclusions, it appears evident that the X-ray outbursts occur  $\approx 8$ days after the periastron passage, marked by the optical orbital period of 111.0 days. In the case of short and sharp X-ray outburst, like 811213-E, the time delay with respect to the optical relative maximum (optical flare 811205-E), at the periastron passage, is just 8 days. This bear witness that Flavia' star was in a `quiescent' state (normal X-ray outburst -- GSG92: Figs. 1). The same for the normal X-ray outbursts reported in Figs. 2, 3, and 8. On the contrary, when Flavia' star is `excited', or `expelling a shell', its luminosity increases much more and the following anomalous X-ray outbursts (GSG92) (Figs. 4, 5, 9, 11a and 11c) or casual X-ray outburst ((GSG92) (Figs. 6, 11b, and 12) are much more intense and with long duration, and start about 8 days after the periastron passage, where the optical flares occur, and develop for many days. The exception is the double peaked X
 -ray outburst of Fig. 11d for which it is reasonable to think that it occurred casually because of a particular excited state of the optical companion.

Thus we can say that the A0535+26/Flavia' star system experiences an optical luminosity enhancement of $\approx 0.02-0.2$ mag, peaked at the periastron passage of the neutron star followed, after $\approx 8$ days, by X-ray outbursts of intensity dependent on the optical activity.
In order to justify quantitatively such a delay we have developed a model that we will describe in the following sections.

%\newpage
\section{Observational properties of the X-ray transient A0535+26/HDE245770}

The observed orbital period $P$ and eccentricity $e$ are

  $$P \simeq 111\,\,{\rm days} \hfill, e = 0.47 \hfill$$
From the Kepler law we have a relation

 \begin{equation}
 P = 2\pi {a^{3/2}}\sqrt {\frac{1}{{G(M + m)}}},
  \label{eq1}
 \end{equation}
  where $M=14\, M_\odot$ and $m=1.4\, M_\odot$ are the masses of the optical and neutron stars (used in this paper), respectively. The large semiaxis of the orbit $a$, and minimal separation of the stars at the periastron $r_{\min}$ from equation (1) are determined as (Landau \& Lifshitz, 1988)

\begin{equation}
 a = \sqrt[3]{{\frac{M}{{{M_{\odot}}}}}}{(\frac{P}{{{P_{\rm Earth}}}})^{2/3}}\sqrt[3]{{1 + \frac{m}{M}}}\approx 1.8 \cdot 10^{13}\, {\rm cm},
  \label{eq2}
 \end{equation}

\begin{equation}
  {r_{\min }} = a(1 - e)\approx 9.6 \cdot 10^{12}\, {\rm cm}.
  \label{eq3}
 \end{equation}

 where $P_{\rm Earth}= 1$ year.
 In quasistatic approximation,  a radius of the Roche lobe around the neutron star $r_{roche}$, at periastron, may be written
 (Eggleton, 1983)

%\begin{equation}
  $${r_{\rm roche}} = {r_{\min }}\frac{{0.49{q^{2/3}}}}{{0.6{q^{2/3}} + \ln (1 + {q^{1/3}})}}$$
\begin{equation}
\qquad \qquad \qquad \approx 0.58\, r_{\min}=5.5 \cdot 10^{12}\, {\rm cm},
  \label{eq4}
 \end{equation}
where $q$ is the mass ratio

 $$q = \frac{m}{M}=0.1 \hfill.$$

\section{Model of the time delay}

Suppose that in the vicinity of the periastron the mass flux $\dot M$ rapidly increases, reaching $\dot M \approx 10^{-7} M_\odot$/year.
We consider here  $\dot M $ as a mass flux trough the accretion disk, entering the neutron star. This mass flux during the flare is much greater than the possible flux during the quiescent phase, being the accretion disk probably temporary (Giovannelli et al., 2007).
Therefore we neglect this hypothetical small flux in further considerations.
 The outer parts of the accretion disk become hotter, giving increase to the optical luminosity (optical flush). Due to large turbulent viscosity, the wave of the large mass flux is propagating to the central object. When this wave reaches the vicinity of the neutron star the X-ray luminosity increases due to appearance of a hot accretion disk region, and due to rapid increase of the luminosity when the accretion flow is  channeled by the magnetic field, and falls onto the magnetic poles (Bisnovatyi-Kogan, 2002; Bisnovatyi-Kogan \& Fridman, 1969).

We identify the time delay $\tau$ between the optical and X-ray flashes, with the time during which the wave of a high mass accretion flux, starting from the outer surface $r_{\rm out}$, reaches the central compact star radius $r_{\rm in}$. The speed of this wave is approximately equal to the radial speed $v_r$ of the matter in the accretion disk, corresponding to this high mass flux. The time delay $\tau$ is calculated below in this model.

 For simplicity, we consider the geometrically thin, optically thick accretion disk without advection, around a compact object.
We suggest, that at each radius, with $\dot M(r,t)$, the accretion disk parameters are the same as in the stationary accretion disk with the same $\dot M$ over the whole disk (rapid relaxation approximation).
 The system of equations describing such disk, in a locally stationary approximation,
 has a following form (Shakura \& Sunyaev, 1973; Bisnovatyi-Kogan, 2002).
 The mass conservation equation is

\begin{equation}
\dot {M}=4\pi rh\rho v_r.
  \label{eq6}
 \end{equation}
The angular momentum equation is

\begin{equation}
\frac{\dot {M}}{4\pi }\frac{dl}{dr}+\frac{d}{dr}(r^2ht_{r\varphi })=0.
  \label{eq7}
 \end{equation}
After integration it reduces to a form

\begin{equation}
r^2ht_{r\varphi } =-\frac{\dot {M}}{4\pi }(l-l_{in} ),
  \label{eq8}
 \end{equation}
where $\ell=\Omega r^2$  is a specific angular momentum, and $ t_{r\phi}=-\alpha P$ is the ($r,\phi$) component of the viscosity tension tensor (Shakura \& Sunyaev, 1973). The other components of this tensor are
assumed to be negligibly small;
$\Omega$ is the Kepler angular velocity $\Omega^2=GM/r^3$. The value of the $l_{in}$ is determined by the angular momentum at the inner boundary of the accretion disk, with zero derivative of the angular velocity. For the accretion into a black hole it corresponds to the Keplerian angular momentum at the last stable orbit, at $r_{\rm in}=3r_g=\frac{6GM}{c^2}$. For the accretion into a non-magnetized, slowly rotating  neutron star $r_{\rm in}$ is close to its radius $R_{\rm ns}$ (Bisnovatyi-Kogan, 2002). In the case of a strongly magnetized star, where the Alfv\'{e}n radius
 $r_A\gg R_{\rm ns}$ a definition of the inner radius and inner angular momentum are less clear, because the matter flows along the field lines to magnetic poles from the Alfv\'{e}n surface, and the disk model failed. In our problem the time delay is determined by a slow radial motion in the outer parts of the accretion disk, where  $l\gg l_{in} $ and therefore the choice of the value of $l_{\rm in}$ is not so  important. For simplicity, we use the value corresponding formally to the black hole.
The local equation of the energy conservation (Shakura \& Sunyaev, 1973) is written as

\begin{equation}
Q^+=Q^-\,\, {\rm (erg/cm^2/s)},
  \label{eq9}
 \end{equation}
where $Q^+$ is the energy production rate by a viscous dissipation, related to the unit of the disk surface,

\begin{equation}
Q^+=ht_{r\varphi } r\frac{d\Omega }{dr},
  \label{eq10}
 \end{equation}
 and $Q^-$ is a radiative flux from the optically thick disk, through the unit of the disk surface

\begin{equation}
Q^-=\frac{2acT^4}{3\tau _0}.
  \label{eq11}
 \end{equation}
 Here $T$ is the temperature, $a$ is the constant of the radiation density, and $\tau_{0}=\kappa \rho h$ is the Thompson optical depth, given by $\tau_{0}=0.4 \rho h$ for a hydrogen composition.
The pressure $P_{\rm tot}$ is determined by a mixture of matter $P_{\rm gas}$ and radiation $P_{\rm rad}$, as

\begin{equation}
P_{\rm tot}=P_{\rm gas}+P_{\rm rad}=\rho{\cal R}T\,+\,{a T^4 \over 3}
  \label{eq12}
 \end{equation}
Here the gas and radiation pressures are given by a standard formulae,
    where ${\cal R}$ is the gas constant.
This system of equations is reduced to a single algebraic non-linear equation for the sound speed $c_s$ (Artemova et al., 1996)

\begin{equation}
  c_s^2 + \frac{{\dot M\kappa r\Omega '{c_s}}}{8 \pi c}f - \sqrt[4]{{-\frac{{{{\dot M}^2 R^4 \kappa }{f^2}r\Omega '{\Omega ^2}}}{{32 \pi^2 a c  \alpha c_s^2}}}} = 0,
  \hfill
  \label{eq15}
 \end{equation}
  where $\Omega'=\frac{d\Omega}{dr}$, and   $f = 1 - \sqrt {\frac{l_{in}}{l}}=1 - \sqrt {\frac{6GM}{rc^2}}$.
Solving this equation we find $c_s$. Radial velocity of the matter in the disk is given by a formula
\begin{equation}
  {v_r} = \frac{{\alpha c_s^2}}{{r\Omega f}}. \hfill
  \label{eq16}
 \end{equation}

The  parameters that we can change in our problem are the viscosity parameter $\alpha$, and the initial mass flux function $\dot M(r)$.
The mass of the neutron star is supposed to be known with sufficient accuracy from the observations.
We suggest that the flush of matter happens when the neutron star passes around the periastron.
The initial distribution of matter flux in the disk is defined as follows

\begin{equation}
\dot{M}(r)=\frac{r_{\sigma} ^{2} }{(r-r_{0} )^{2} +r_{\sigma} ^{2} }{\dot M_0}
\end{equation}
Here $r(t)$ define the Lagrangian radius of the disk matter, $r_0=1.1\cdot 10^{11}$\,cm is a point of the maximum mass flux over the disk at the initial time, and $r_{\sigma}=3.2\cdot 10^{10}\,$cm is a parameter, determining the
 characteristic width the high mass flux region. We take, for the most intensive outburst, $\dot M_0 = 10^{-7}
 M_\odot$/year as a maximum value of $\dot M$ (Persi et al., 1979),  $\dot M_0 = 3\cdot 10^{-8}
 M_\odot$/year and $\dot M_0 = 10^{-8}
 M_\odot$/year for less powerful outbursts (de Loore et al., 1984).
The disk is formed at the radius $r_0$, at which the velocity, $v_{flow}$, of the matter flowing out from the star is of the order of the Keplerian velocity $v_K=\sqrt{GM/r_0}$. For the above mentioned $r_0$ the reasonable value of the outflow velocity is $v_{flow}\approx 410$ km s$^{-1}$.
 Furthermore, knowing the local solution for the radial velocity of
matter $v_r$, for the structure of the disk at some point $r$ in time $ t
$ we  find the time evolution of the initial distribution of
matter, in the time interval $ dt $ by the formula

\begin{equation}
 r(t+dt)=r(t)-v(r)dt
 \end{equation}

The total mass falling onto the neutron star during the flare in this model is determined by the initial $\dot M$ distribution (eq. 14). In the locally static
approximation a surface density of the accretion disk at the radius $r$ depends on the local value of $\dot M$, as well as of  $M$ and $\alpha$. The physical conditions in the accretion disk at radius $r\sim r_0$ are characterized by  a gas pressure and Krammers opacity. The surface density $\Sigma$ in this region written as (Bisnovatyi-Kogan, 2010), taking the system parameters

\begin{equation}
\Sigma=1.5\cdot 10^4 \frac{{\dot m}^{7/10}m^{1/5}}{\alpha^{4/5}x^{3/4}}, \,\,\,  m=\frac{M}{M_\odot}=1.4,\,\,\, {\dot m}= \frac{{\dot M}c^2}{L_c},
 \end{equation}
  $$L_c=\frac{4\pi c GM}{\kappa_T}\approx 1.3\cdot 10^{38}\frac{M}{M_\odot} {\rm \frac{erg}{c}}\approx 1.8\cdot 10^{38}{\rm \frac{erg}{c}}  $$
  $$x=\frac{rc^2}{GM}=\frac{r}{1.5\cdot 10^5\,{\rm cm}}\frac{M_\odot}{M}\approx \frac{r}{2.1\cdot 10^5\,{\rm cm}}  $$

Introducing $\dot M_{-7}=\frac{\dot M}{10^{-7}M_\odot/{\rm year}}$, we obtain

\begin{equation}
\dot m\approx 33\dot M_{-7},\,\,\, \Sigma \approx 10\frac{\dot M_{-7}^{7/10}}{\alpha^{4/5}(r/r_0)^{3/4}}\,{\rm g/cm^2},
 \end{equation}

where $ r_0=1.1\cdot 10^{11}$ cm. The mass falling onto the neutron star during the flare $M_{flare}$ is obtained by the integration of $\Sigma$ over the disk, with the distribution of the $\dot M$ from eq.~14. We have

\begin{equation}
 M_{flare}=2\pi\int_0^{\infty}\Sigma rdr=20\pi\frac{\dot M_{-7}^{7/10}}{\alpha^{4/5}}{r_\sigma}^2
 \end{equation}
$$\times\int_0^\infty \frac{y^{1/4}dy}{(y-1)^2+0.085} {\rm g},\,\,\, y=\frac{r}{r_0},\,\, \left(\frac{r_\sigma}{r_0}\right)^2=0.085. $$

The integral in eq. 18 is converging rapidly in both limits, therefore we may integrate from zero to infinity making only small error. Taking into account $r_\sigma=3.2\cdot 10^{10}$, and making the integration

 $$\int_0^\infty \frac{y^{1/4}dy}{(y-1)^2+0.085}\approx 10,$$

we obtain the value of $M_{flare}$ in the form

  \begin{equation}
M_{flare}\approx 3\cdot 10^{-10}M_\odot \frac{\dot M_{-7}^{7/10}}{\alpha^{4/5}}.
 \end{equation}

Note, that during the luminous phases the radiation is coming from the disk radius which is much smaller than the initial one, and where the opacity is determined mainly by the Thomson scattering. This is taken into account in the subsequent calculations.
Knowing how the matter in the disk is moving, we can calculate  the variation of the luminosity of the disk with time, in the optical and  X-ray ranges. The luminosity of this object will consist of two components: emission from the disk and emission from hot spots, which are formed on the neutron star by a matter falling onto the magnetic poles (Fig. 13). The integrated luminosity of the disk is determined by the following formula

\begin{equation}
I=\int _{r_{in} }^{r_{out} }dI(r),\,\,\,
dI(r)=2\pi r \int _{\lambda _{1} }^{\lambda _{2} }\frac{2hc^{2} }{\lambda ^{5} } \frac{d\lambda }{\exp (\frac{hc}{\lambda kT} )-1}
\label{eq16a}
\end{equation}

where $dI(r)$ is a radiation coming from an elementary ring at the radius  $r$,  and thickness $dr$, in the wavelength range from $\lambda_1$ to $\lambda_2$, calculated by the formula of blackbody radiation at temperature $T$.
To calculate the integrated luminosity in the optical band,  the integral (eq. 16)) was calculated  in the range 300--700 nm, and for the luminosity in the X-rays band it was done  in the range  2--10 keV.

The second component of the radiation comes from the hot
spots. Hot spot is formed at the magnetic poles of the neutron star
due to infall of the  matter from the disk along the magnetic field lines.
The magnetic field lines, form a column of matter over the
poles of the neutron star. The angular size of the bottom of the column on the neutron star surface is approximately defined by  the  expression $\sin ^{2} \theta =\frac{r_{ns} }{r_{A} }$ (Baan \& Treves, 1973),
where $r_{ns}$ is the radius of the neutron star, and $r_{A}$  is the radius of the Alfv\'{e}nic surface  which is
determined by the following expression (Lamb, Pethick \& Pines, 1973)

%\begin{equation}
$$r_{A} =\left(\frac{B^{2} r_{ns}^{6} }{6\dot{M}\sqrt{2GM_{ns} } } \right)^{2/7}$$
\begin{equation}
\qquad \qquad \qquad
=5.8\cdot 10^7\left(\frac{B_{12}^{2} r_{6}^{6} }{\dot{M}_{7}\sqrt{GM_{\odot} } } \right)^{2/7} {\rm cm}.
\end{equation}

For our system the equatorial magnetic field on the neutron star surface is taken equal to  $B=4\cdot 10^{12}\,$ G (Terada et al., 2006), the  maximum accretion rate $\dot M_0 \approx 10^{-7} M_\odot$/year (Persi et al., 1979), and the radius of the neutron star $r_{ns}=10\,$km.
Knowing the size of the hot spots, it is possible to estimate its effective temperature.

\begin{equation}
T=\left(\frac{\eta \dot{M}c^2}{S\sigma} \right)^{1/4} =2.4\cdot 10^8 \left(\frac{\dot{M}_{7}}{S_{10} } \right)^{1/4} K
\end{equation}

where $S=\pi r_{ns}^2\left(\frac{r_{ns}}{r_{A}}\right)$ is the surface of the hot spot,
$\eta\approx \frac{r_{g} }{r_{ns} } =0.3$ is the efficiency of a conversion of the kinetic energy of a falling matter into a radiation.
 Knowing the effective temperature and the surface of the hot spot it is possible to calculate the radiation power in
the optical and X-ray spectral bands, similar to what was done for the accretion disk.

%%%%%%%%%%%%%%%%%%% FIGURE 13 %%%%%%%%%%%%%%%%%
\begin{figure}[hbp] %[!ht] %[!hbp]%[!ht]%[!hbp]%[h][h!]
\begin{center}
\includegraphics[width=9cm]{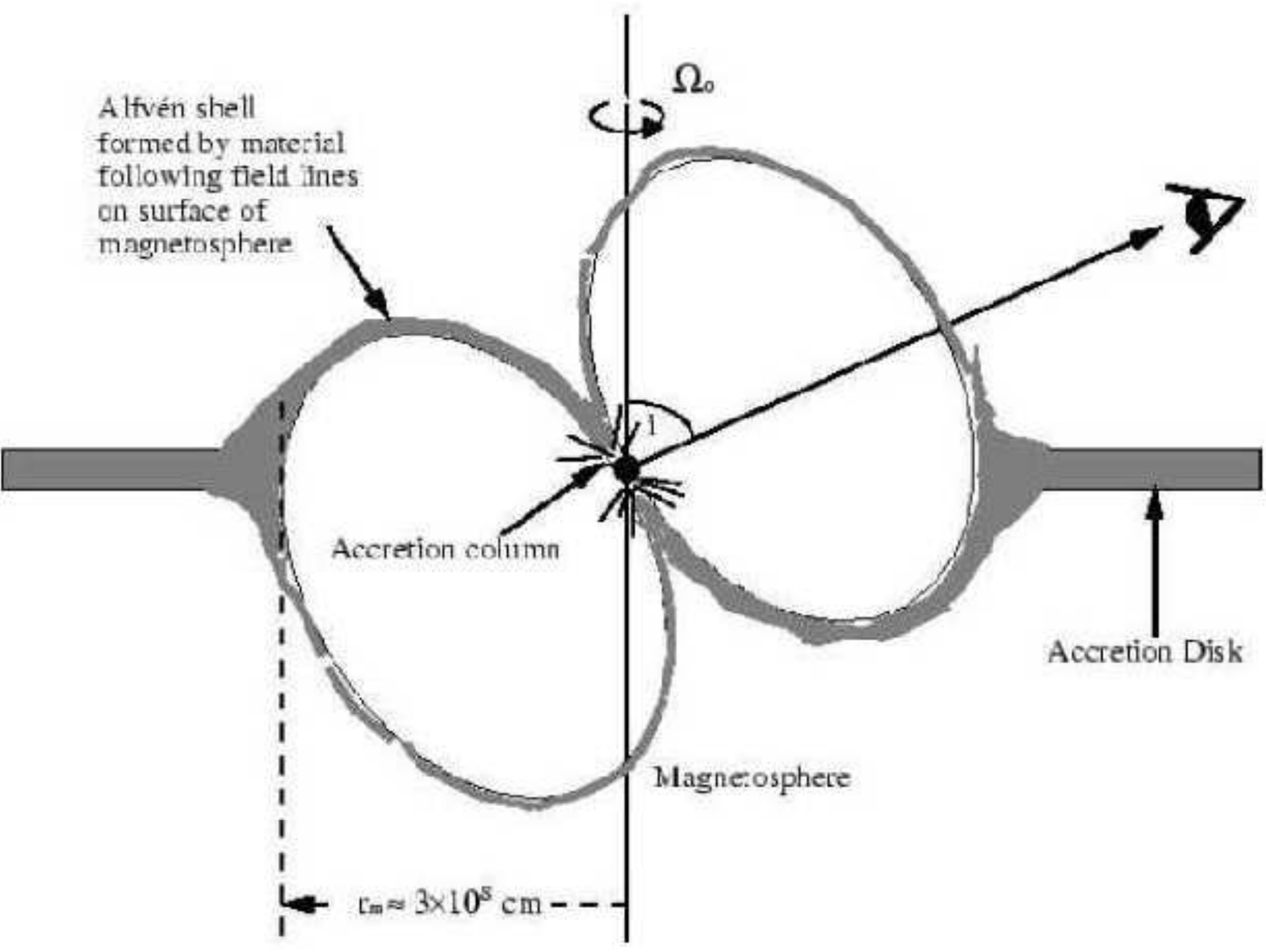}\\
\caption{Sketch of the accretion flow in a disk being picked up by a strong neutron star magnetic field.} \label{lc13}
\end{center}
\end{figure}
%%%%%%%%%%%%%%%%%%%%%%%%%%%%%%%%%%%%%%%%%%%%%%

\section{Results}

We have calculated the time dependence of the radiation power of the object in two ranges.  In the optical  (300--700 nm) and X-rays (2--10 keV) ranges.
The results for the light curves of the system in these two bands are given in Fig. 14. Upper, middle, and lower panels of Fig. 14 refer to different values of $\dot M_0$ and $\alpha$, as indicated in the figures. The bolometric luminosity is also reported in each panel.

%%%%%%%%%%%%%%%%%%% FIGURE 14 %%%%%%%%%%%%%%%%%
\begin{figure}[hbp] %[!ht] %[!hbp]%[!ht]%[!hbp]%[h][h!]
\begin{center}
\includegraphics[width=6cm]{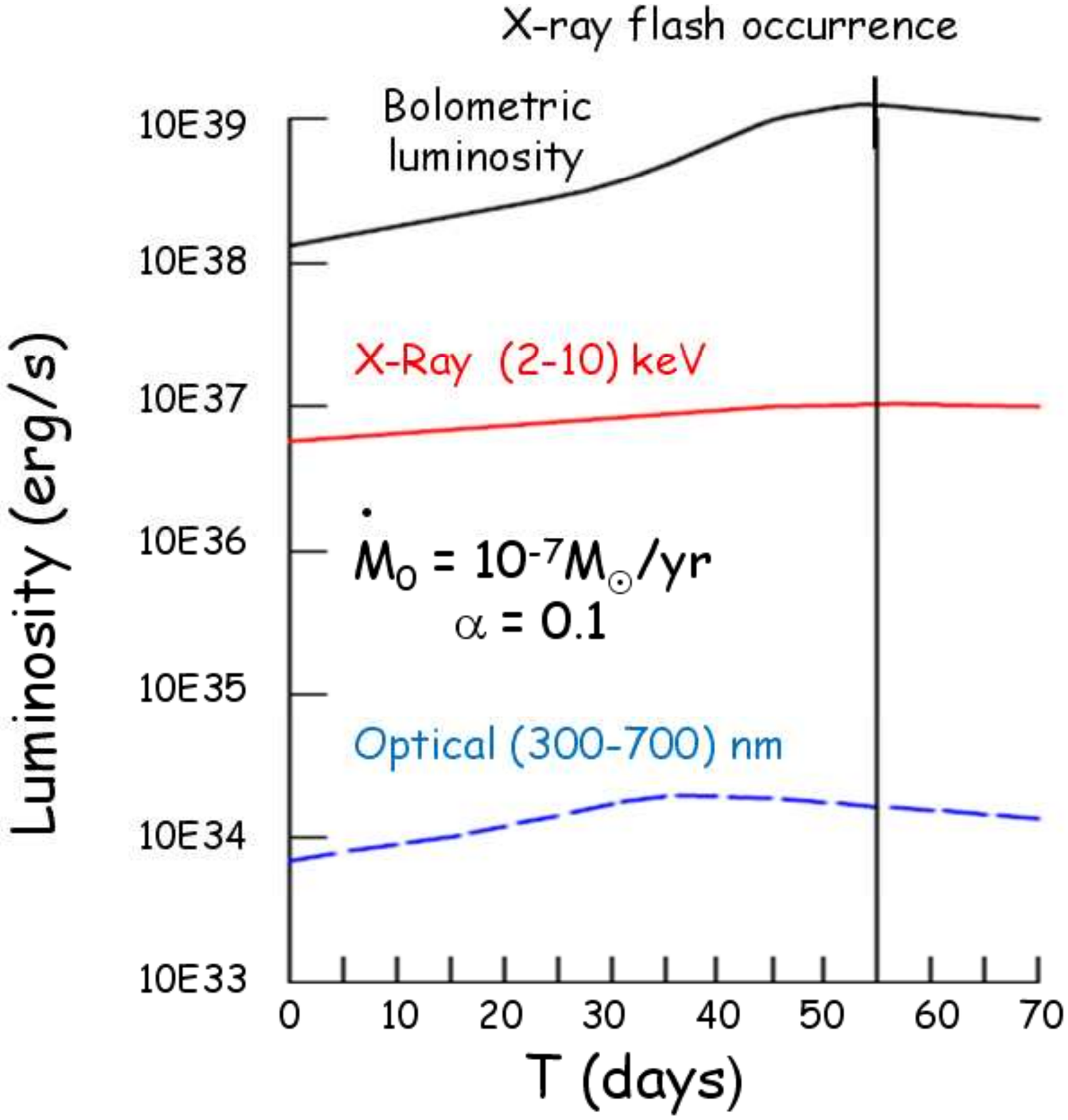}
%\vspace{-1.5cm}
\includegraphics[width=6cm]{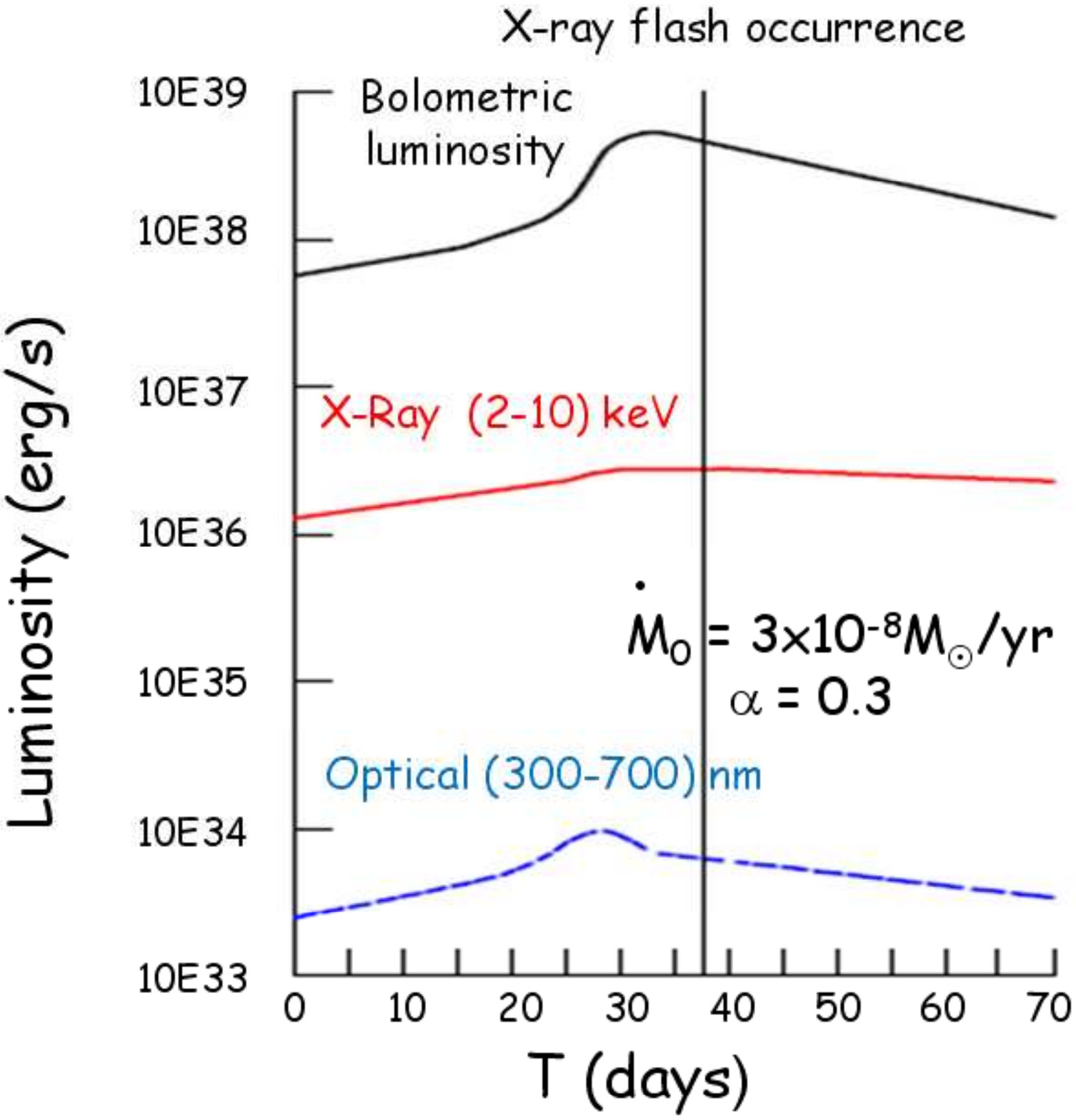}\\
\vspace{-1.5cm}
\includegraphics[width=6cm]{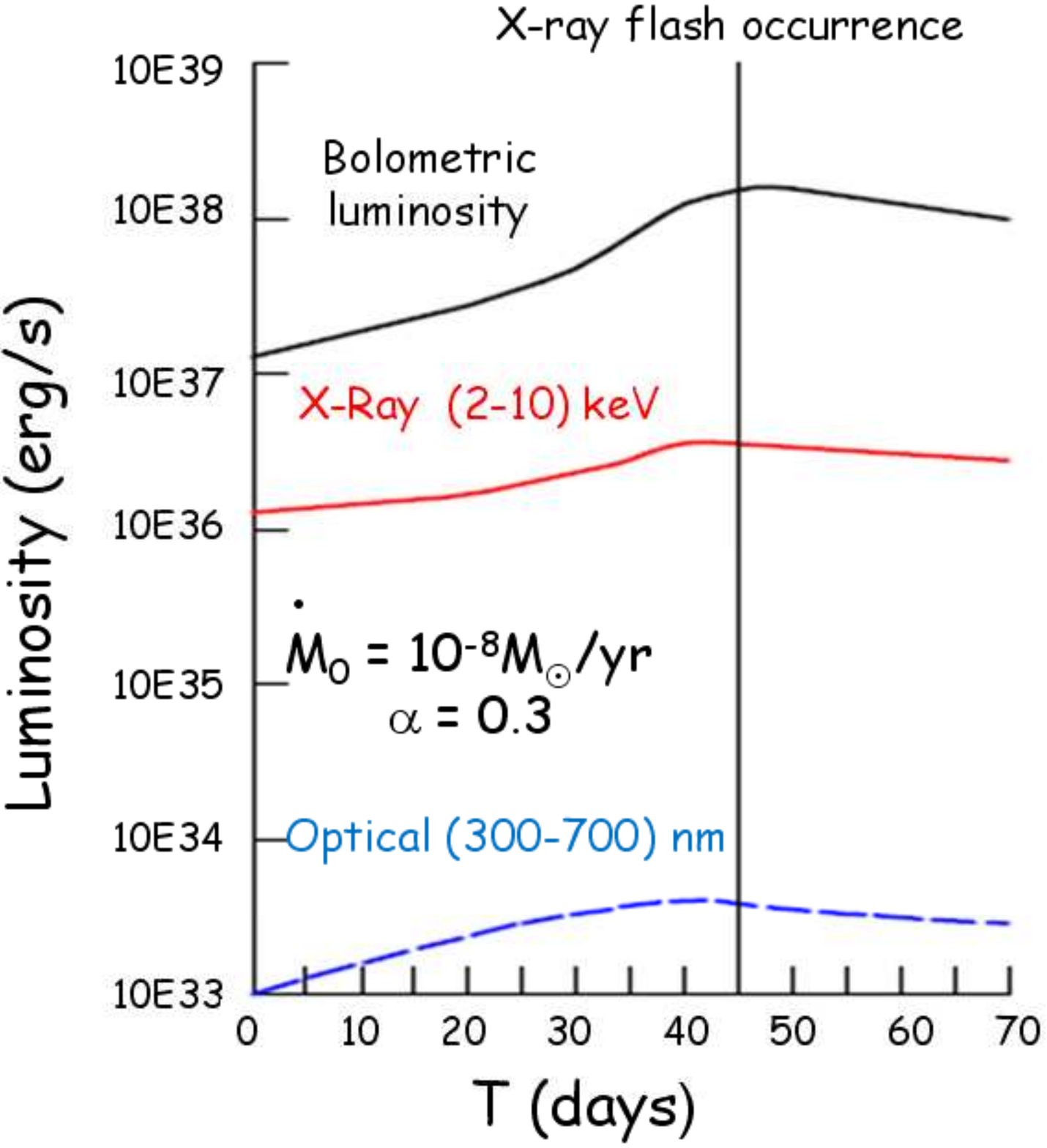}\\
\vspace{-1cm}
\caption{The time dependence of the: i) bolometric luminosity of the object (black line), ii) X-rays in the band 2-10 keV (red line),  and iii) optical in the band 300-700 nm (broken blue line). Black vertical line marks the X-ray (2-10 keV) flash occurrence,  about 8 days after the maximum of the optical curve. $\dot M$ is taken from (14), with the following $\dot M_0$.
Upper-left panel: $\dot M_0=10^{-7}\,M_\odot$/year, $\alpha=0.1$. Upper-right panel: $\dot M_0=3\cdot 10^{-8}\,M_\odot$/year, $\alpha=0.3$. Lower panel: $\dot M_0=10^{-8}\,M_\odot$/year, $\alpha=0.3$.}  \label{lc14}
\end{center}
\end{figure}
%%%%%%%%%%%%%%%%%%%%%%%%%%%%%%%%%%%%%%%%%%%%%%

%%%%%%%%%%%%%%%%%%% FIGURE 15 %%%%%%%%%%%%%%%%%
\begin{figure}[hbp] %[!ht] %[!hbp]%[!ht]%[!hbp]%[h][h!]
\begin{center}
\includegraphics[width=6cm]{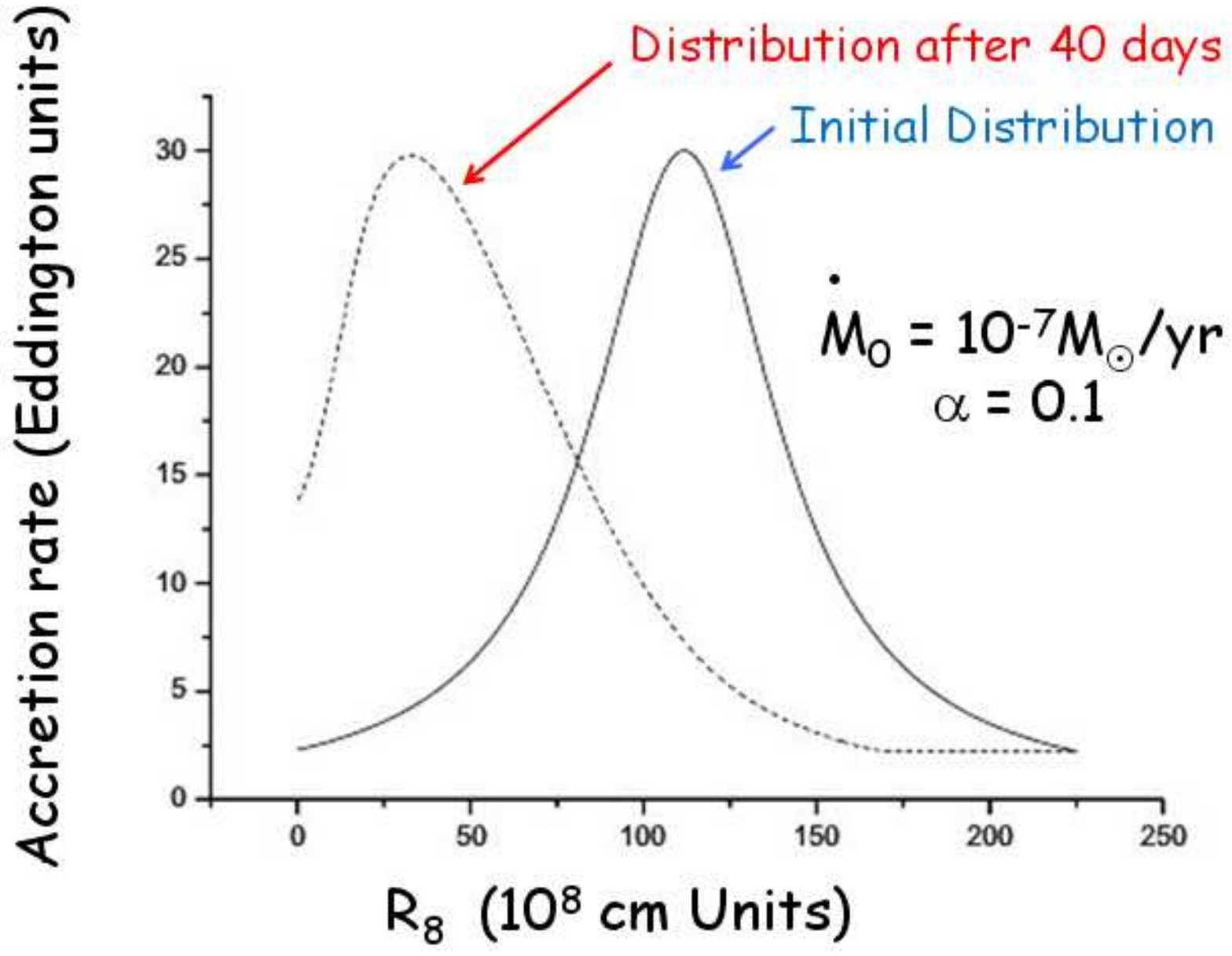}
%\vspace{-3cm}
\includegraphics[width=6cm]{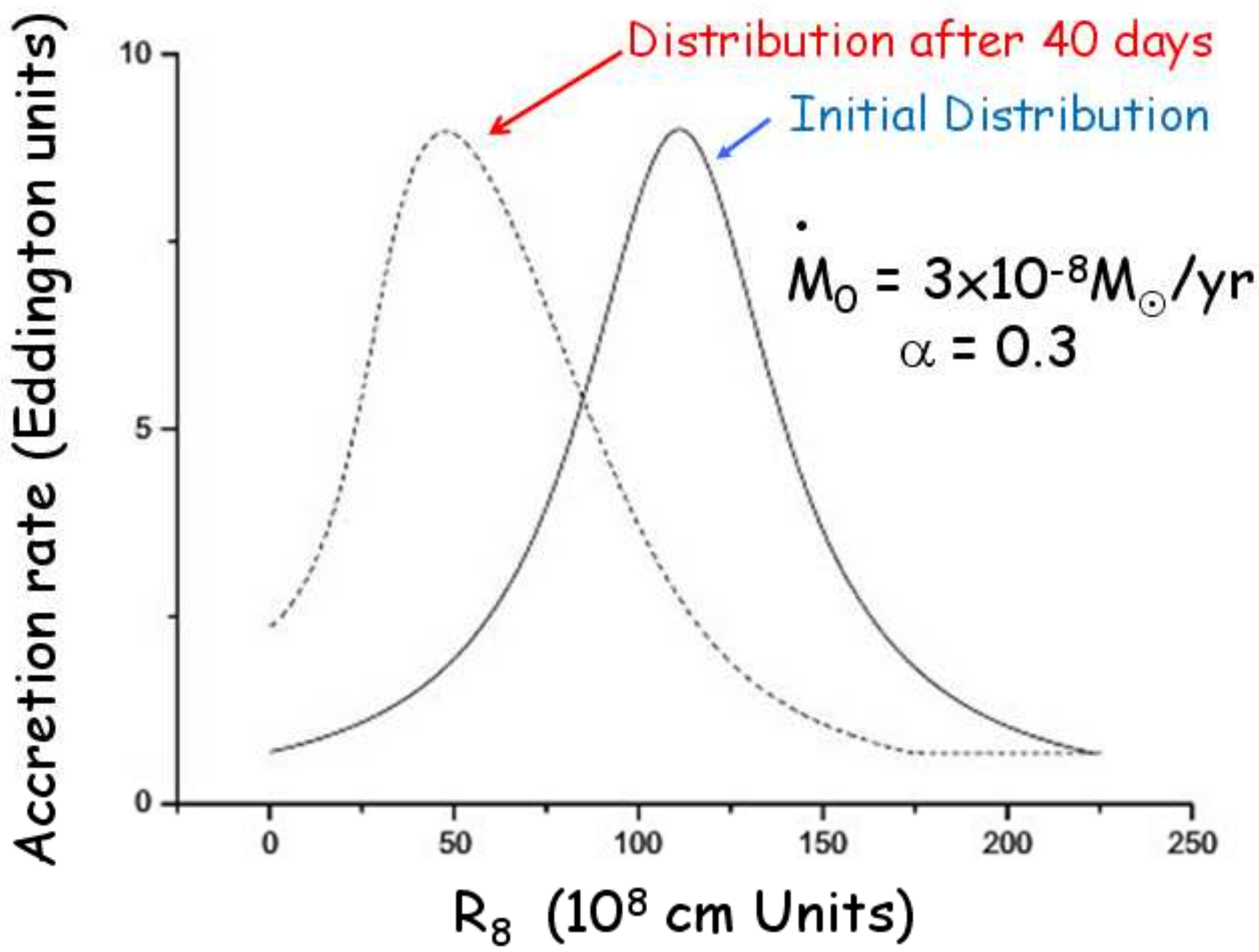}\\
\vspace{-3cm}
\includegraphics[width=6cm]{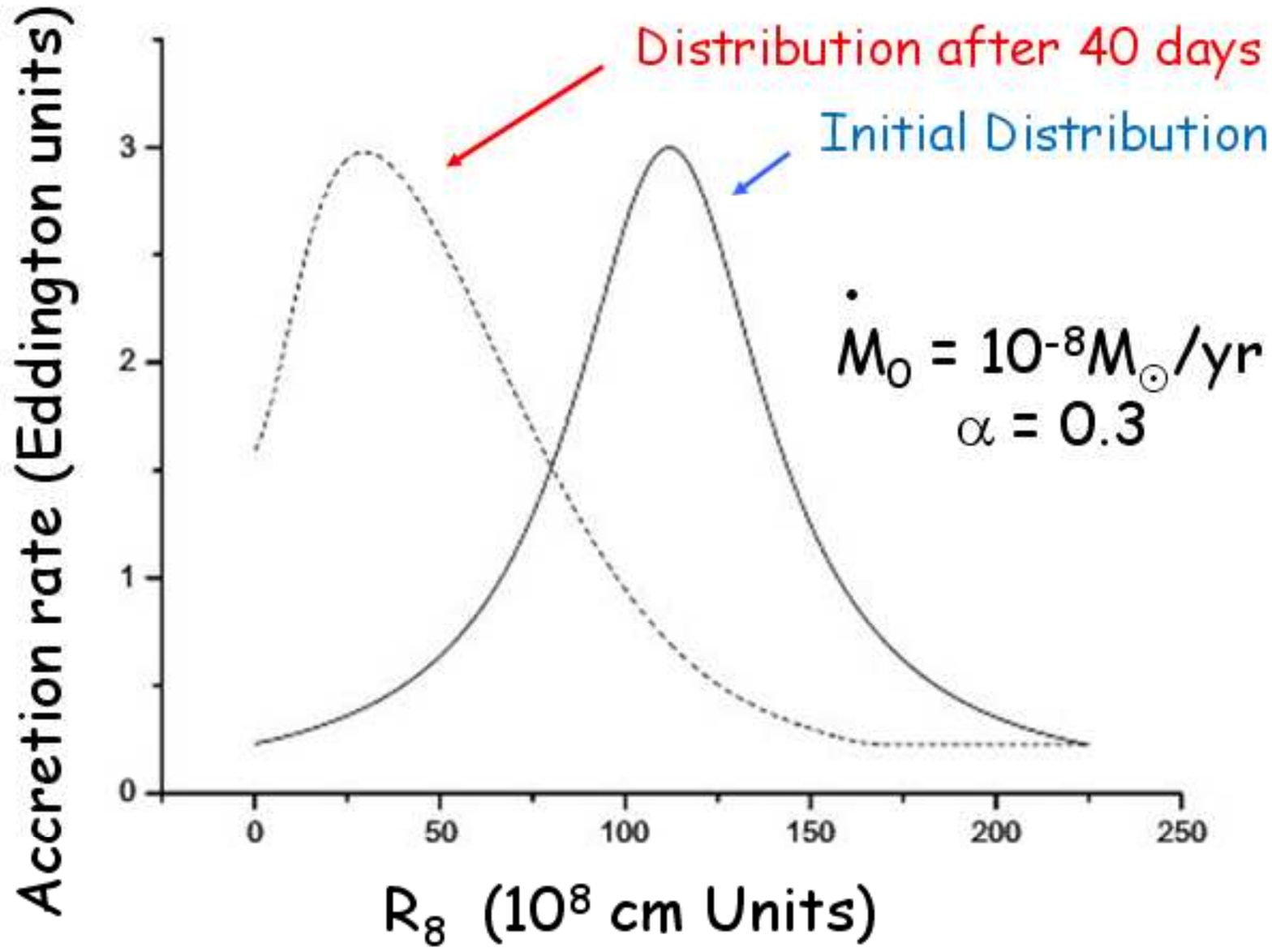}\\
\vskip-1cm
\caption{The radial dependence of the  accretion rate in the disk.
Accretion rate is in the units of the  Eddington  accretion rate, and R$_8$ is in $10^8$ cm units.
 Solid line  shows the initial distribution and, the broken line shows the distribution of accretion rate on the disk after 40 days of evolution.
 $\dot M$ is taken from (14), with the following $\dot M_0$. Upper-left panel: $\dot M_0=10^{-7}\,M_\odot$/year, $\alpha=0.1$. Upper-right panel:  $\dot M_0=3\cdot 10^{-8}\,M_\odot$/year, $\alpha=0.3$. Lower panel:  $\dot M_0=10^{-8}\,M_\odot$/year, $\alpha=0.3$.} \label{lc15}
\end{center}
\end{figure}
%%%%%%%%%%%%%%%%%%%%%%%%%%%%%%%%%%%%%%%%%%%%%%

It shows that the maxima of the time dependence of the light curves in  different bands are shifted relative to each other.
 The maximum radiation in the X-ray range is reached later than the maximum of radiation at optical wavelengths.
The main parameter of the problem that we have been able to vary is the Shakura-Sunyaev viscosity parameter $\alpha$ for the accretion disk.
The time delay between the peaks in the optical and X-ray is directly dependent on this parameter.
We have found that  the time delay between the maxima of the X-ray  and optical emissions  that we need for explaining of the observational data (for the object A0535+26/HDE245770 it is equal to 8 days) is reached at a value of the  viscosity parameter $\alpha \approx 0.1$ for the powerful flash with $\dot M_0 = 10^{-7} M_\odot$/year (Fig. 14, upper panel); and at a value of the  viscosity parameter $\alpha \approx 0.3$ for the flashes with $\dot M_0 = 3\cdot 10^{-8} M_\odot$/year, and $\dot M_0 = 10^{-8} M_\odot$/year (Fig. 14 middle and lower panels, respectively). The maximum and minimum values of $\dot M_0$ were derived from the papers by Persi et al. (1979) and de Loore et al. (1984), respectively. The mass of the gas falling onto the neutron star during the flare is found from eq. 19

\begin{equation}
 M_{flare} \approx 1.9\cdot 10^{-9}M_\odot\quad {\rm for}\,\,\, \dot M_{-7}=1,\,\,\,\alpha=0.1,
 \end{equation}
$$M_{flare} \approx 3.4\cdot 10^{-10}M_\odot\quad {\rm for}\,\,\, \dot M_{-7}=0.3,\,\,\,\alpha=0.3,$$
$$M_{flare} \approx 1.6\cdot 10^{-10}M_\odot\quad {\rm for}\,\,\, \dot M_{-7}=0.1,\,\,\,\alpha=0.3.$$

This delay can be explained quite simply. At the beginning the disk was formed due to strong increase of the mass flux when the neutron star approaches the periastron of the orbit. The main emission initially comes from outside, relatively cold regions of the disk, and
 the maximum luminosity comes just at the optical range.
The matter flux rate in the inner parts of the disk is small at this stage and the contribution of the whole disk to the X-ray luminosity is small. The mass flows from the outer to the inner regions as a result of the evolution of the disk.

 The contribution of the outer regions in  the optical radiation decreases, while the contribution of the radiation from  inner hot disk in the X-ray band begins to grow.  The radiation from the hot spots at the poles of the neutron star is added, which  makes a significant contribution to the radiation in the X-rays.

Figure 15 shows the radial dependence of the  accretion rate in the disk.
Accretion rate is in the units of the  Eddington  accretion rate. Solid line shows the initial distribution and the broken line shows the distribution of accretion rate on the disk after 40 days of evolution. Upper, middle, and lower panels of Fig. 15 refer to different values of $\dot M_0$ and $\alpha$, as indicated in the figure caption.
It is also evident from Fig. 15, that the graphs of the time dependence of the radiation power in X-rays is asymmetrical with respect to the maximum. The increase of radiation power occurs more rapidly than the decline. This is because the initial symmetric distribution of matter in the disk during its evolution becomes asymmetrical. This is consistent with the observed asymmetry in the X-ray light curves of A 0535+26.

Similar behavior may be expected in other high-mass transient systems with highly eccentric orbits. The time delay between optical luminosity enhancement -- occurring at the periastron -- and X-ray flashes with asymmetric light curves is expected to be similar to that measured in A0535+26/HDE245770.

Note that, according to our model, during the falling flow onto the accreting star, the increase of the
luminosity should always start from the low energy part of the spectra, and the maximum in the high energy band follows that in the low energy.
This fact can be clearly seen in Fig. 8 where the maximum of the X-ray outburst in the range 2--10 keV (SAS III measures) is reached $\sim 8$ days before the maximum in the range 28--40 keV (Prognoz 6 satellite -- Signe II experiment measures).

This is contrary to that occurring during the ejection of the hot gas from a star or active galactic nucleus (AGN), where the increasing of luminosity starts from the high energy side, and the maximum in the low energy part happens after.

\section{Conclusions}

 The ephemeris used in this paper -- JD$_{\rm opt-outb}$ = JD$_0$(2,444,944) $\pm$ n(111.0 $\pm$ 0.4) days, derived from the orbital period of the system P$_{\rm orb} = 111.0 \pm 0.4$ days and from the optical flare of December 5, 1981 (811205-E), followed by the X-ray flare of December 13, 1981 (811213-E), allowed us to demonstrate that the A0535+26/Flavia' star system scans the neutron star passage at the periastron with this periodicity. This passage is roughly 8 days before any kind of X-ray outbursts, but that of June--August 2010.

It appears evident that also the very strong X-ray outbursts, named casual outbursts by GSG92, follow the periastron passage of about 8 days. Indeed, even in the case when a shell is ready to be expelled from the O9.7 IIIe star, the event is triggered at the periastron, because of gravitational corkscrew rule. A huge amount of matter is going onto the temporary accretion disk around the neutron star, producing an enhancement in the optical luminosity. This matter slowly flows onto the neutron star producing the 8-day-delayed strong X-ray outburst.
This evidence is astonishingly confirmed looking at the light curve of A0535+26 (Rosenberg et al., 1975), when the X-ray source was detected by the Ariel V satellite: the detection of the outburst started on April 13, 1975 (the first Ariel V measure), and the periastron passage -- accordingly with our ephemeris -- occurred on  March 30, 1975 (JD 2,442,502), $\sim 14$ days  before the beginning of the X-ray outburst, exactly 22 orbital periods before 811205-E (see Fig. 12).

We have constructed a  quantitative model of this event, basing on a nonstationary accretion disk behavior, connected with a high ellipticity of the orbital motion. The observed time delay is related to the motion of a high-mass flux region from the outer boundary of the NS Roche lobe to the Alfv\'{e}n surface due to the action of the $\alpha$--viscosity.
For bright outbursts the 8 days delay happens for $\alpha = 0.1$.

Our model could be valid also in the cases of AGNs. Indeed, Nandra et al. (1998) found a delay of $\sim$ 4 days between UV and X-ray emissions in NGC 7469; Maoz, Edelson \& Nandra (2000) found a delay of $\sim$ 100 days between optical and X-ray emissions in the Seyfert galaxy NGC 3516; Marshall, Ryle \& Miller (2008) found a delay of $\sim$ 15 days between optical and X-ray emissions in Mkr 509, and Doroshenko et al. (2009) found a delay of $\approx$ 10 days between R, I and X-ray luminosities in the Seyfert galaxy 3C 120.

It is interesting to note that a short qualitative explanation of these lags in AGNs, very similar to our quantitative model, was suggested by Marshall, Ryle \& Miller (2008).

\section{Acknowledgements}
We would like to thank the anonymous referee for his/her very useful suggestions that rendered this paper much more clear, Mr Francesco Reale and Mr Massimo Frutti for their help in the preparation of several figures.
This research made use of NASA's Astrophysics Data System.


\begin{thebibliography}{}


\bibitem{abbn}
Artemova, I. V., Bisnovatyi-Kogan, G. S., Bjoernsson, G., Novikov, I. D.: 1996, ApJ  456, 119

\bibitem{bat} Baan W.A. Treves A.: 1973, A\&A, 22, 421

\bibitem{1} Bartolini, C., Guarnieri, A., Piccioni, A., Giangrande, A., Giovannelli, F.: 1978, IAU Circ. No. 3167

\bibitem{2} Bartolini, C., Bianco, G., Guarnieri, A., Piccioni, A.,
Giovannelli, F.: 1983, Hvar Obs. Bull. 7(1), 159.

\bibitem{bk} Bisnovatyi-Kogan, G. S.: 2010,  Stellar Physics. Vol. 2. Stellar
evolution and stability. Berlin Heidelberg: Springer (2d Edition)

\bibitem{bf} Bisnovatyi-Kogan, G. S., Fridman A.M.: Astron. Zh., 1969, 46, 721

\bibitem{3} Burger, M., van Dessel, E.L., Giovannelli, F., Sabau-Graziati, L., Bartolini, C. et al.: 1996, in
{\it Multifrequency Behaviour of High Energy Cosmic Sources}, F. Giovannelli \& L. Sabau-Graziati (eds.), Mem. SAIt. 67, 365
%Guarnieri, A.; Piccioni, A.

\bibitem{4}	Caballero, I., Kretschmar, P., Santangelo, A., Staubert, R., Klochkov, D., et al.: 2007, A\&A 465, L21
%Camero, A.; Ferrigno, C.; Finger, M. H.; Kreykenbohm, I.; McBride, V. A.; Pottschmidt, K.; Rothschild, R. E.; Sch�nherr, G.; Segreto, A.; Suchy, S.; Wilms, J.; Wilson, C. A.

\bibitem{5} Caballero, I., Lebrun, F., Rodriguez, J., Soldi, S., Mattana, F., et al.: 2010a, ATel. 2496
%Santangelo, A.; Wilms, J.; Kreykenbohm, I.; Kretschmar, P.; Ferrigno, C.; Pottschmidt, K.; Rothschild, R.

\bibitem{6} Caballero, I., Santangelo, A., Pottschmidt, K., Klochkov, D., Rodriguez, J., et al.: 2010b, ATel. 2541
%Wilms, J.; Kreykenbohm, I.; Kretschmar, P.; Ferrigno, C.; Rothschild, R.; Suchy, S.

\bibitem{7} Caballero, I., Pottschmidt, K., Barrag\'{a}n, L., Ferrigno, C., Klochkov, D., et al.: 2010c, Talk at CRSF Meeting,
T\"{u}bingen 2010

\bibitem{8} Caballero, I., Kretschmar, P., Pottschmidt, K., Santangelo, A., Wilms, J., et al.: 2010d, AIPC 1248, 147
%Kreykenbohm, I.; Ferrigno, C.; Suchy, S.; Rothschild, R.; Finger, M.; Postnov, K.; McBride, V.; Domingo, A.; Sch�nherr, G.; Klochkov, D.; Staubert, R.; Camero-Arranz, A.

\bibitem{9} Caballero, I., Pottschmidt, K., Santangelo, A., Barrag\'{a}n, L., Klochkov, D., et al.: 2011, arXiv:1107.3417
%Ferrigno, C.; Rodriguez, J.; Kretschmar, P.; Suchy, S.; Marcu, D. M.; Mueller, D.; Wilms, J.; Kreykenbohm, I.; Rothschild, R. E.; Staubert, R.; Finger, M. H.; Camero-Arranz, A.; Makishima, K.; Mihara, T.; Nakajima, M.; Enoto, T.; Iwakiri, W.; Terada, Y.

\bibitem{10}	Camero-Arranz, A., Finger, M.H., Wilson-Hodge, C.A., Jenke, P., Steele, I., et al.: 2012, ApJ 754, 20
%Coe, M. J.; Gutierrez-Soto, J.; Kretschmar, P.; Caballero, I.; Yan, J.; Rodr�guez, J.; Suso, J.; Case, G.; Cherry, M. L.; Guiriec, S.; McBride, V. A.

\bibitem{coe} Coe, M.J., Carpenter, G.F., Engel, A.R., Quenby, J.J.: 1975, Nature  256, 630

 \bibitem{11}  Coe, M.J., Reig, P., McBride, V.A., Galache, J.L., Fabregat, J.: 2006, MNRAS 368, 447

\bibitem{12} Doroshenko, V.T., Sergeev, S.G., Efimov, Yu.S., Klimanov, S.A., Nazarov, S.V.: 2009, Astron. Lett. 35, 361

\bibitem{egl}  Eggleton P.P. 1983, ApJ 268, 368

\bibitem{13} Finger, M.H., Cominsky, L.R., Wilson, R.B., Harmon, B.A.,
Fishman, G.J.: 1994, in {\it The Evolution of X-Ray Binaries},
S.S. Holt \& C.S. Day (eds.), AIP Conf. Proc., 308, 459

\bibitem{14} Finger, M.H., Wilson, R.B., Harmon, B.A.: 1996, ApJ 459, 288

\bibitem{15} Giangrande, A., Giovannelli, F., Bartolini, C., Guarnieri, A., Piccioni, A.: 1980, A\&AS 40, 289.

\bibitem{gio84b}
Giovannelli, F., Ferrari-Toniolo, M., Persi, P., Golynskaya, I.M., Kurt, V.G., et al.: 1984, in
X-Ray Astronomy '84, M. Oda and R. Giacconi (eds.), Institute of Space and Astronautical
Science, Tokyo, p. 205.

\bibitem{16}
Giovannelli, F., Ferrari-Toniolo, M., Persi, P., Golynskaya,
I.M., Kurt, V.G., et al.: 1985, in {\it Multifrequency Behaviour of Galactic Accreting Sources}, Proceedings of the 1984 Frascati Workshop, Franco Giovannelli (ed.). Frascati: CNR, Istituto di Astrofisica, Ed. Scientifiche SIDEREA, Roma,  p. 284

\bibitem{gz90}	
Giovannelli, F.; Ziolkowski, J.: 1990, AcA 40, 95

\bibitem{gl92}	
Giovannelli, F.; Sabau-Graziati, L.: 1992, SSRv 59, 1 (GSG92)

\bibitem{gbrs}
Giovannelli, F.; Bernabei, S.; Rossi, C.; Sabau-Graziati, L.: 2007, A\&A  475, 651G

\bibitem{gl92a}	
Giovannelli, F., Gualandi, R., Sabau-Graziati, L.: 2010, ATel. 2497

\bibitem{fg}  Giovannelli F,  Sabau-Graziati L, 2011, Acta Polytechnica 51, No. 2., 21

\bibitem{17} Gnedin, Y.N., Zaitseva, G.V., Larionov, V.M., Lyutyi, V.M., Khozov, G.V., Sheffer, E.K.:
1988, Sov. Astron. 32(6), 624

 \bibitem{17a} Guarnieri, A., Bartolini, C., Piccioni, A., Giovannelli, F.: 1985, in {\it Multifrequency Behaviour of Galactic Accreting Sources}, Proceedings of the 1984 Frascati Workshop, Franco Giovannelli (ed.). Frascati: CNR, Istituto di Astrofisica, Ed. Scientifiche SIDEREA, Roma,  p. 310

 \bibitem{18} Hutchings, J.B.: 1984, PASP 96, 312

 \bibitem{19} Janot-Pacheco, E., Motch, C., Mouchet, M.: 1987, A\&A 177, 91

\bibitem{20} Joss, P.C., Rappaport, S.A.: 1984, Ann. Re. A\&A 22, 537

\bibitem{kal}
Kaluzienski, L.J., Holt, S.S., Boldt, E.A. Serlemitsos, P.J.: 1975,  Nature 256, 633

\bibitem{21} 	van Kerkwijk, M.H., van Paradijs, J., Zuiderwijk, E.J.: 1995, A\&A 303, 497

\bibitem{lpp} Lamb F.K., Pethick C.J., Pines D.: 1973, ApJ 184, 271

\bibitem{ll}  Landau L.D., Lifshitz E.M. 1988, Vol. 1 Mechanics. Nauka. Moscow

\bibitem{dlgd}
de Loore, C., Giovannelli, F., van Dessel, E.L., Bartolini, C., Burger, M., et al. 1984, A\&A 141, 279
%De Loore, C.; Giovannelli, F.; van Dessel, E. L.; Bartolini, C.; Burger, M.; Ferrari-Toniolo, M.; Giangrande, A.; %Guarnieri, A.; Hellings, P.; Hensberge, H.; Persi, P.; Piccioni, A.; van Diest, H.

\bibitem{22} Lyuty, V.M., Zaitseva, G.V.: 2000, Astron. Lett. 26, 9

\bibitem{23} Maoz, D., Edelson, R., Nandra, K.: 2000, AJ 119, 119

\bibitem{24} Margon, B., Nelson, J., Chanan, G., Bowyer, S., Thorstensen, J.R.: 1977, ApJ 216, 811

\bibitem{25} Marshall, K., Ryle, W.T., Miller, H.R.: 2008, ApJ 677, 880

 \bibitem{26} de Martino, D., Vittone, A., Giovannelli, F., Ciatti, F., Margoni, R., et al.: 1985, in {\it Multifrequency Behaviour of Galactic Accreting Sources}, Proceedings of the 1984 Frascati Workshop, Franco Giovannelli (ed.). Frascati: CNR, Istituto di Astrofisica, Ed. Scientifiche SIDEREA, Roma,  p. 326

 \bibitem{26a} Motch, C., Stella, L., Janot-Pacheco, E., Mouchet, M.: 1991, ApJ 369, 490

\bibitem{nag82}
Nagase, F., Hayakawa, S., Kunieda, H., Makino, F., Masai, K., et al.: 1982, ApJ 263, 814

\bibitem{27} Nagase, F., Hayakawa, S., Tsuneo, K., Sato, N., Ikegami, T., et al.: 1984, PASJ 36, 667
%Kawai, N.; Makishima, K.; Matsuoka, M.; Mitani, K.; Murakami, T.; Oda, M.; Ohashi, T.; Tanaka, Y.

\bibitem{28} Nandra, K., Clavel, J., Edelson, R.A., George, I.M., Malkan, M.A., et al.: 1998, ApJ 505, 594
%Mushotzky, R. F.; Peterson, B. M.; Turner, T. J.

\bibitem{29}	Persi, P., Ferrari-Toniolo, M., Spada, G., Conti, G., di Benedetto, P. et al.:
%Tanzi, E. G.; Tarenghi, M.
1979, MNRAS, 187, 293

 \bibitem{30} Priedhorsky, W.C., Terrell, J.: 1983, Nature 303, 681

\bibitem{rick75}
Ricketts, M.J., Turner, M.J.L., Page, C.G., Pounds, K.A.: 1975, Nature, 256, 631

\bibitem{ariel}
Rosenberg, F.D., Eyles, C.J., Skinner, G.K., Willmore, A.P.: 1975, Nature 256, 628

\bibitem{31} R\"{o}ssiger, S,: 1978, IBVS 1395

\bibitem{32} Sembay, S., Schwartz, R.A., Orwig, L.E., Dennis, B.R., Davies, S.R.: 1990, ApJ 351, 675

\bibitem{ss} Shakura, N.I., Sunyaev, R.A.: 1973, A\&A 24, 337

\bibitem{33} Terada, Y., Mihara, T., Nakajima, M., Suzuki, M., Isobe, N., et al.: 2006, ApJ 648, L139
%Makishima, K.; Takahashi, H.; Enoto, T.; Kokubun, M.; Kitaguchi, T.; Naik, S.; Dotani, T.; Nagase, F.; Tanaka, T.; Watanabe, S.; Kitamoto, S.; Sudoh, K.; Yoshida, A.; Nakagawa, Y.; Sugita, S.; Kohmura, T.; Kotani, T.; Yonetoku, D.; Angelini, L.; Cottam, J.; Mukai, K.; Kelley, R.; Soong, Y.; Bautz, M.; Kissel, S.; Doty, J.

\bibitem{34} Thorsett, S.E., Arzoumanian, Z., McKinnon, M.M., Taylor, J.H.: 1993, ApJ 405, L29

\bibitem{35} Violes, F., Niel, M., Bui-van, A., Vedrenne, G., Chambon, G., et al.: 1982, ApJ 263, 320

 \bibitem{} Yan, J., Li, H., Liu, Q.: 2012, ApJ 744, 37


\end{thebibliography}
 \end{document}